\documentclass[fleqn,usenatbib]{mnras}

\usepackage{newtxtext,newtxmath}

\usepackage{orcidlink}

\usepackage[T1]{fontenc}

\DeclareRobustCommand{\VAN}[3]{#2}
\let\VANthebibliography\thebibliography
\def\thebibliography{\DeclareRobustCommand{\VAN}[3]{##3}\VANthebibliography}

\usepackage{graphicx}	
\usepackage{amsmath}	
\usepackage{multirow}
\usepackage[section]{placeins}
\usepackage[version=4]{mhchem}
\usepackage{lipsum}
\usepackage{xspace}
\usepackage{anyfontsize}
\usepackage[normalem]{ulem}

\definecolor{darkgreen}{rgb}{0.0, 0.7, 0.0}
\newcommand{\St}{\ensuremath{\mathrm{St}}\xspace}

\title[Water snowline: I. Effects on mid-IR spectra]{Around the water snowline: I. Effects of dynamical and chemical interplay on mid-infrared water spectra of protoplanetary discs}

\author[T. Kaeufer et al.]{
Till Kaeufer$^{\orcidlink{0000-0001-8240-978X}, 1}$\thanks{E-mail: t.kaeufer@exeter.ac.uk},
Sebastiaan Krijt$^{\orcidlink{0000-0002-3291-6887},1}$,
Edwin A. Bergin$^{\orcidlink{0000-0003-4179-6394},2}$,
Andrea Banzatti$^{\orcidlink{0000-0003-4335-0900},3}$,
and Joe Williams$^{\orcidlink{0009-0008-8176-1974},1}$
\\
$^{1}$ Department of Physics and Astronomy, University of Exeter, Exeter EX4 4QL, UK\\
$^{2}$ Department of Astronomy, University of Michigan, 1085 S. University Ave., Ann Arbor, MI 48109, USA \\
$^{3}$ Department of Physics, Texas State University, 749 N Comanche Street, San Marcos, TX 78666, USA
}

\date{Accepted 2026 September 15. Received 2026 September 8; in original form 2026 July 24}

\pubyear{\the\year{}}

\begin{document}
\label{firstpage}
\pagerange{\pageref{firstpage}--\pageref{lastpage}}
\maketitle

\begin{abstract}

Protoplanetary discs’ infrared water spectra (e.g. as observed by JWST) encode information about the underlying water distribution and physical conditions. In this work, we examine and quantify the imprint of various dynamical processes (operating alone or in unison) on water abundances and emerging spectra using a new 2D model called Molecular emission Affected by Gas, Pebble, and Ice Evolution (MAGPIE). This model combines diffusion and advection for water (and associated species) and dust grains with aerodynamically decoupled pebbles, and photodissociation and a gas-phase water formation prescription. In general, we find that transport processes have a major impact on the inner disc’s water content and can (re)plenish the layer above the ice reservoir which in static models is water-poor due to photodissociation. In particular, adding diffusion to a chemical model can result in strong changes in the spectrum (e.g. a flux increases of colder compared to warmer water lines) that recent work has associated with pebble drift. Adding pebble drift dramatically increases the inner disc water abundance but, surprisingly, does not readily lead to a further increase of cold water tracers as (1) efficient outward transport of water vapour is inhibited by a conveyor belt effect of incoming pebbles, and (2) the increase in hot water line fluxes overshadows smaller increases of colder lines. We conclude that dynamical processes have a significant but complex impact on water spectra. 
Future work should explore a range in stellar (e.g. UV irradiation) and disc properties to comprehensively understand the observed diversity of water emission.

\end{abstract}




\section{Introduction \label{sec:intro}}

Mid-infrared (mid-IR) observations probe the innermost regions ($<10\,\rm au$) of protoplanetary discs. Planets commonly form in these regions \citep{Morbidelli2012}. Therefore, the planet compositions are influenced by the conditions in these disc regions \citep[e.g.][]{Oberg2011,Bitsch2019,Molliere2022}.
Water is of special interest due to its importance for the formation process of planets via its impact on dynamics and solid accretion \citep[e.g.][]{Ros2013,Schoonenberg2017,Drazkowska2023} and the impact on planet habitability \citep[e.g.][]{krijt2023}.

Observations of different water lines probe different water reservoirs. Typically, less than a few percent of the total water vapour reservoir is visible in mid- and far-IR observations \citep{Bosman2022,Leemker2025}. This means that these observations only probe a thin layer in the disc's surface, while most water mass is concealed closer to the midplane due to dust opacities \citep{houge_smuggling_2025}. However, mid-IR spectra taken by the Spitzer space telescope frequently observed a high forest of water lines \cite[][]{Carr2008,Salyk2008}. While these observations were initially modelled with a single temperature component \citep[e.g.][]{Carr2011,Salyk2011}, recent observations from the Mid-InfraRed Instrument (MIRI) on board the James Webb Space Telescope (JWST), benefitting from increased spectral resolution and sensitivity compared to Spitzer, clearly showed deviations from single temperature emission, supporting gradients first emerged by combining ground-based spectra \citep{Banzatti2023a}. Multiple temperature components \citep[e.g.][]{Temmink2024}, radial power laws \citep[e.g.][]{Kaeufer2026}, tapered power laws \citep[e.g.][]{Romero-Mirza2024}, or even more complex profiles \citep[][]{Temmink2025} are now used to analyse water spectra as observed with JWST/MIRI.

Analyses of water spectra show that different discs are dominated by different emission temperatures \citep{Banzatti2023a,Banzatti2023}. \cite{Romero-Mirza2024,Banzatti2025} and Romero-Mirza et al. submitted show for samples of $8-40$ discs that compact discs exhibit a cold water ($T\sim200\,\rm K$) excess compared to extended discs. A cold water excess means that the ratio between water lines with lower and higher upper level energies is enhanced. Specifically, two lines with upper level energies around $1500\,\rm K$ are enhanced compared to a water line with an upper level energy of $3600\,\rm K$. Alternatively, studies have tried to link variations in cold water excess to the positions of the innermost gap \citep{Krijt2025}.
Both the radial disc size and radial location and depth of the innermost resolved gap are proposed to be related to the magnitude of ongoing pebble drift reaching the disc's snowline \citep{Kalyaan2021,Kalyaan2023}. The underlying idea is that pebbles drift inwards over time, releasing their icy mantles into the gas-phase when crossing the species' iceline leading to an increase in emission close to the iceline temperature of $\sim120-180\,\rm K$ \citep[][]{Lodders2003}.
While for other species this process can be complicated by entrapment of volatiles inside water ice \citep[e.g.][]{Barnun1985,Bergner2024,Williams2025}, water as the main icy species with the highest sublimation temperature is commonly thought to be released at the water iceline, the so-called snowline. 

Several efforts have been made to model the drift of icy pebbles and release of volatiles in detail.
\cite{Cuzzi2004} show that drifting pebbles can increase the water vapour concentration inside the snowline by factors of up to $10-100$ \citep[see also][who reached enhancement factors of up to $10$]{Ciesla2006}. Other studies highlight the dependence on grain size for the exact location where grains desorb their volatiles \citep{Piso2015}, the process of water delivery to planetesimal embryos \citep{Sato2016}, and linking planetary compositions to their formation history \citep{Booth2017,Schneider2021}.

However, linking the well-studied process of pebble drift and ice sublimation to JWST/MIRI spectra is challenging because the above mentioned models only include the radial dimension. As a result, to connect the two, studies have either resorted to simplified analytical arguments \citep[e.g.][]{Romero-Mirza2024,Krijt2025}, built in parametrized elevated abundances in 2D models \citep{Vlasblom2025}, or extended 1D dynamical models to 1+1D \citep{Sellek2025,houge_smuggling_2025}.
These works highlight the strong spectral impact of drift on mid-IR spectra through the desorption of species that are directly observable \citep[][]{Sellek2025}, through an increase of dust opacity hiding more disc material \citep{houge_smuggling_2025}, and through the change of elemental abundances which affects the inner disc's chemistry \citep{Sellek2025b}.

In this paper, we address these problems by creating a 2D (radial and vertical) disc model, called Molecular emission Affected by Gas, Pebble, and Ice Evolution (MAGPIE). The goal is to directly link the transport processes that are dominating in the midplane to higher disc regions that are frequently observed. This model includes dynamical processes (diffusion and advection) for water vapour, water ice, small dust grains, other gas-phase species, and pebbles as well as a simple chemical description to describe photodissociation and gas-phase water formation. We showcase the influence of all the modelled processes on the water density structure and simulate time-dependent mid-IR spectra and diagnostic diagrams that allow for a comparison with JWST/MIRI observations.

This paper is organised as follows. We introduce the model in Sect.~\ref{sec:method}. Section~\ref{sec:results_density} and Section~\ref{sec:results_spec} compare the evolution of the water concentrations and observables of models that include different processes. The impact of these processes and the wider implications are discussed in Sect.~\ref{sec:discus} before the paper is concluded (Sect.~\ref{sec:conclusion}).

\section{Methods}
\label{sec:method}

\begin{figure*}
    \centering
    \includegraphics[width=1.0\linewidth]{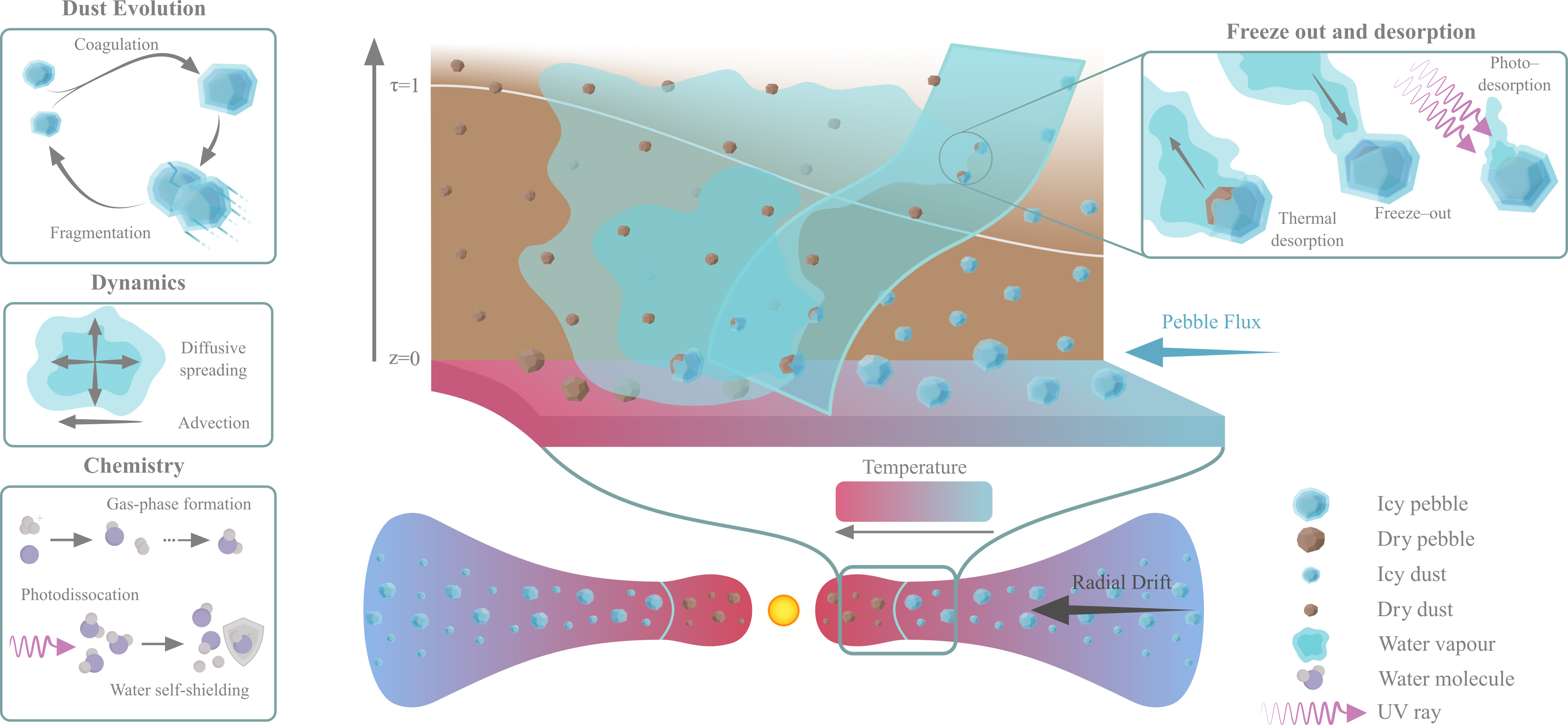}
    \caption{Schematic representation of a protoplanetary disc (bottom) with a zoom-in (centre) onto the water snowline. The schematic shows how in-drifting icy pebbles cross the water snowline, sublimate their ice, and enrich the water gas-phase abundance. The four boxes show all processes modelled by MAGPIE. Dust evolution entails coagulation of small  dust grains into pebbles and fragmentation of these pebbles. Dynamics covers diffusion and advection, which includes pebble drift as well. The chemical processes modelled by MAGPIE include gas-phase formation of water via \ce{H3+} and photodissociation accounting for shielding by \ce{H2O} and \ce{OH}. Additionally, we model thermal and photodesorption of ice and freeze out onto dust grains and pebbles. Lastly, the background  of the main panel indicates the underlying mid-IR opacity structure which is solved time dependently to simulate JWST/MIRI observations. \citep[Inspired by Fig.~1 of][]{houge_smuggling_2025}}
    \label{fig:model_sketch}
\end{figure*}

This section introduces the Molecular emission Affected by Gas, Pebble, and Ice Evolution (MAGPIE) model, which is used to examine the influence of dynamics and chemistry on the gas-phase water abundance. All processes that influence the water abundance and are included in the model are highlighted in Fig.~\ref{fig:model_sketch}. After listing the disc setup (Section~\ref{sec:struc_general}), we explain the implementation of transport processes (Section~\ref{sec:transp}), ice-gas interactions (Section~\ref{sec:ice}), chemistry (Section~\ref{sec:chem}), pebbles (Section~\ref{sec:pebbles}), and the numerical details (Section~\ref{sec:numerics}). This model follows many ideas introduced by \cite{Ciesla2009,Krijt2018,Krijt2020}, but consists of a new implementation optimised for the inner disc region around the water snow line, an implementation of the most important chemical processes of water, and a fast description of pebble transport. Additionally, the disc densities can be translated to simulated JWST/MIRI spectra (Section~\ref{sec:sim_obs}).

\subsection{Disc setup \label{sec:struc_general}}
\subsubsection{Density structure \label{sec:dens_struc}}

\begin{table}
    \centering
    \caption{Parameters of the model listed with their fiducial value.}
    \label{tab:model_parameters}
    \begin{tabular}{l p{4cm} l} \hline \hline
          & \\[-1.9ex]
          Symbol &  Explanation  & Value  \\ \hline
          & Structure \\
          $M_{\rm disc} \, [\mathrm{M_\odot}]$ & Gas mass  & $0.02$\\      
          $M_{\rm dust} \, [\mathrm{M_\odot}]$ & Total solid mass & $0.01M_{\rm disc}$\\ 
          $r_\mathrm{c} \, [\mathrm{au}]$ &Characteristic radius & $30$\\ 
          $p$ & Power law index of the surface density & $1.0$\\          
          $M_{*} \, [\mathrm{M_\odot}]$ & Stellar mass & $0.7$\\  
          $H_{0} \, [\mathrm{au}]$ & Scale height at $r_0=100\,\rm au$ & $10.0$\\  
          $\beta$ & Flaring index & $1.15$\\  
          
          \hline
          & Temperature \\
          $T_{0}\, [\mathrm{K}]$& Midplane temperature at $1\,\rm au$ & $150$ \\
          $q$& Temperature power law index & $0.5$ \\
          $\phi_\mathrm{T}$& Atmospheric temperature increase compared to midplane & $2$ \\
          
          $\delta_{\mathrm{T}}$& Vertical temperature profile & $4.0$ \\
          $z_{\rm q} \, [\mathrm{H_g}]$& Lower limit of atmosphere & $4.0$ \\
          \hline
          & Evolution \\
          $\alpha$& Turbulence & $10^{-3}$ \\
          $f_\mathrm{c}$& Simulation timestep (as fraction of smallest timescale) & $0.1$ \\   
          \hline
          & Dust and pebbles \\
          $s_{\bullet}\, \rm [cm]$ & Small dust grain size & $10^{-5}$ \\
          $\rho_{\bullet}\, \rm [g /cm^3]$ & Dust/pebble grain internal density & $1.6$ \\
          $v_{\rm frag}\,[\mathrm{m/s}]$& Pebble fragmentation velocity & $10.0$\\     
          \hline
          & Ice \\
          $F_{\rm UV}\, \rm [cm^{-2}s^{-1}]$ & Incident vertical UV flux & $10^{8}$ \\
          $Y\, [\mathrm{photon^{-1}}]$& Photodesorption yield & $10^{-2}$\\     
          $N_\mathrm{s}\, [\mathrm{~cm^{-2}}]$ & Adsorption site density & $10^{15}$\\       
          $f_{\mathrm{w}}$ & Stickiness of the ice & 0.5 \\
          $\mathcal{E}/k_\mathrm{B} \,[\rm K]$ & \ce{H2O} binding energy & $5770$\\
          \hline
          & Chemistry \\
          $t_{\rm UV,0} [\mathrm{yr}]$ & Photodissociation timescale & $40.0$ \\
          $\zeta_{\rm CR} [\mathrm{s^{-1}}]$ & Cosmic ray ionisation rate & $1.7\times 10^{-17}$ \\
          $\epsilon_{\rm H_2O}$ & Abundance of \ce{H2O} & $10^{-4}$ \\
          $\epsilon_{\rm CO}$ & Abundance of \ce{CO} & $10^{-4}$ \\
          $\epsilon_{\rm H_2}$ & Abundance of \ce{H2} & $0.5$ \\
          $\epsilon_{\rm O_{tot}}$ & Total oxygen abundance  & $3.2\times 10^{-4}$ \\
          \hline
          & Observations \\
          $d [\mathrm{pc}]$ & Distance to object & $140$ \\
          $i [\mathrm{^\circ}]$ & Inclination & $0.0$ \\
          \hline

    \end{tabular}
\end{table}

We assume a gas surface density distribution ($\Sigma_{\mathrm{g}}$) that follows a radial ($r$) tapered power law:

\begin{align}
    \Sigma_\mathrm{g}(r)=\Sigma_\mathrm{c}\left(\frac{r}{r_\mathrm{c}}\right)^{-p} \exp{\left[-\left(\frac{r}{r_\mathrm{c}}\right)^{2-p}\right]}. \label{eq:surface_dens}
\end{align}

This power law is fixed at the critical radius ($r_\mathrm{c}$) and corresponding critical surface density ($\Sigma_\mathrm{c}$) and depends on the surface density exponent ($p$). The values for all parameters are provided in Table~\ref{tab:model_parameters}. The critical surface density can be related to the total disc mass ($M_{\rm disc}$) using 

\begin{align}
    \Sigma_\mathrm{c}=(2-p)\frac{M_{\rm disc}}{2\pi r_\mathrm{c}^2}.
\end{align}

We assume a parametrised gas scale height ($H_\mathrm{g}$) that follows a power law 
dependent on a scale height $H_0$ at radius $r_0$ and the power law index $\beta$:

\begin{align}
    H_\mathrm{g}(r)=H_0 \left(\frac{r}{r_0}\right)^{-\beta}.
\end{align}

Using this scale height, the gas surface densities can be translated to local densities $\rho_\mathrm{g}$ at any given radius and height over the disc midplane ($z$) with:

\begin{align}
    \rho_\mathrm{g}(r,z)=\frac{\Sigma_\mathrm{g}(r)}{\sqrt{2\pi}H_{\mathrm{g}}}\exp{\left[ -\frac{1}{2}\left(\frac{z}{H_{\mathrm{g}}}\right)^{2} \right]} \label{eq:vert_distr}.
\end{align}

\begin{figure}
    \centering
    \includegraphics[width=\linewidth]{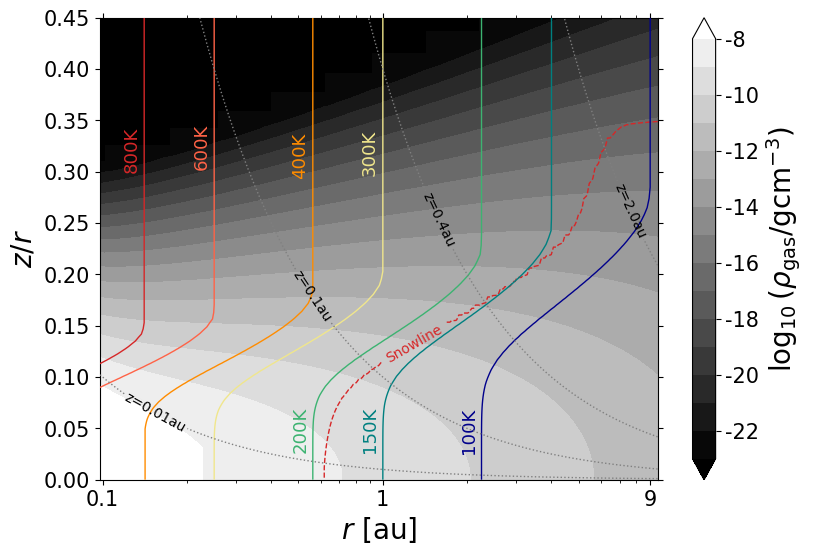}
    \caption{Gas density (colour map) and temperature structure (coloured contours) of the fiducial model. The red dashed snowline marks the transition between the majority of water present in ice or vapour. The dotted lines denote equal disc heights.}
    \label{fig:setup}
\end{figure}

The gas density structure of the model examined in this paper can be seen in Fig.~\ref{fig:setup}. We note that the spatial grid in radius and height (details in Sect.~\ref{sec:num_model_setup}) undersamples small $z/r$ values at small radii (compare to dotted lines in Fig.~\ref{fig:setup}). Therefore, after calculating $\rho_\mathrm{g}(r,z)$ at every grid point, the surface densities obtained by vertically integrating said densities are rescaled to adhere to Eq.~\ref{eq:surface_dens}.

Using a gas-to-dust ratio ($100$ for our fiducial model), the gas density is translated into dust densities $\rho_\mathrm{d}$.

Lastly, the Keplerian frequency $\Omega$ at any given radius is calculated using the star mass ($M_\star$) and the gravitational constant $G$:
\begin{align}
    \Omega(r)=\sqrt{\frac{G M_\star}{r^3}}.
\end{align}

\subsubsection{Temperature structure \label{sec:temp}}

In this study, we fix the temperature structure during the model's evolution and assume that the dust and gas temperature are the same.
In reality, the disc temperature structure is influenced by many factors, some of which will be time dependent. For example, the chemistry affects the disc gas temperature and will cause it to deviate significantly from the dust temperature \citep[][]{Woitke2009}. Additionally, changing the dust density structure or latent heat released by sublimating ices change the radiative transfer solution, leading to significant evolutions in a disc's temperature \citep[][]{Wang2025}. While fixing the temperature structure is not realistic, it is commonly done in disc evolution models in 1D and 2D \citep[e.g.][]{Krijt2018,Stammler2022}. The main concern is the computational cost to solve the radiative transfer equation for every time step. Therefore, this approach is limited so far to individual studies \citep[e.g.][]{Robinson2024}.  

We assume a fixed temperature distribution (Fig.~\ref{fig:setup}). The temperature is governed by the midplane ($T_{\rm mid}$) and disc atmosphere temperature ($T_{\rm atm}$) where the atmosphere is defined as $z_{\rm atm}=z_\mathrm{q} \,H_\mathrm{g}$ in terms of factors ($z_{\mathrm{q}}$) of the radius dependent scale heights. Following previous studies \citep[e.g.][]{Dartois2003,Rosenfeld2013}, we model the transition between the two values as

\begin{align}
    T(r,z) =T_{\rm mid}(r) +\left(T_{\rm atm}(r)-T_{\rm mid}(r)\right)\left[\sin\left({\frac{\pi z}{2 z_\mathrm{q} H_\mathrm{g}(r)}}\right)\right]^{2\delta}
\end{align}
with the strength of the transition determined by $\delta$.

The midplane temperature is defined as a power law dependent on the temperature at $1\,\rm au$ ($T_0$) and the power law index ($q$): $T_{\rm mid}= T_0 \left(r\right)^{-q}$. The parametrised temperature in the disc's atmosphere is a constant factor ($\phi_\mathrm{T}$) higher than the midplane temperature ($T_{\rm atm}(r) = \phi_\mathrm{T} T_{\rm mid}(r)$). This 2D temperature structure mimics more realistic temperature profiles determined via \ce{CO} observation \citep[e.g.][]{Dutrey2017,Galloway2025}. 

We assume that any material immediately adjusts to the cell temperature when entering a given grid cell, because the typical timescale for dust grains in protoplanetary discs to adjust to changing radiation field conditions is in the order of seconds to hours \citep{Chiang1997} which is much shorter than the time steps considered in this study. 

We note that the used temperature profile mimics a passively heated disc. Accretion heating can significantly alter the temperature structure \citep[e.g.][]{Calahan2026}. We refer to Sect.~\ref{sec:possible_solutions} for a discussion on how changes to the temperature structure affect our results.

\subsection{Diffusion \& Advection \label{sec:transp}}

The core goal of this model is to dynamically evolve water vapour in the disc setup highlighted in Section~\ref{sec:struc_general}. As shown in the bottom right box of Fig.~\ref{fig:model_sketch}, we assume a viscous disc and solve the advection-diffusion equation for all relevant quantities:

\begin{equation}
    \begin{split}
 \dfrac{ \partial C_i }{\partial t} = &\overbrace{\frac{1}{r \rho_\mathrm{g}} \dfrac{\partial}{\partial r }\left( r \rho_\mathrm{g} D_i \dfrac{\partial C_i}{\partial r} \right)}^{\mathrm{Radial\, \,diffusion}} &- \overbrace{\frac{1}{r \rho_\mathrm{g}} \dfrac{\partial}{\partial r }\left( r \nu_{r,i} \rho_\mathrm{g} C_i \right)}^{\mathrm{Radial \,\,advection}}\\
 &+  \underbrace{\frac{1}{\rho_\mathrm{g}} \dfrac{\partial}{\partial z }\left( \rho_\mathrm{g} D_i \dfrac{\partial C_i}{\partial z}  \right)}_{\mathrm{Vertical \, \, diffusion}} &-  \underbrace{\frac{1}{\rho_\mathrm{g}} \dfrac{\partial}{\partial z }\left(\nu_{z,i} \rho_\mathrm{g} C_i \right)}_{\mathrm{Vertical \, \, advection}} .
\end{split}
    \label{eq:advect_diff}
\end{equation}

In this equation, $C_i$ denotes the concentration ($\rho_i/\rho_\mathrm{g}$) of a species $i$, with $D_i$ being the corresponding diffusion coefficient and $\nu_{r,i}$ and $\nu_{z,i}$ describing the radial and vertical advection velocity. We evolve water vapour ($C_\mathrm{H_2O}$), small dust grains ($C_\mathrm{d}$), ice on small dust grains ($C_{\rm ice}$), OH vapour ($C_{\rm OH}$), and oxygen vapour ($C_{\rm O}$; more details in Section~\ref{sec:chem}).

\cite{Shakura1973} parametrise the turbulent viscosity in a gas disc as $\nu_\mathrm{T}=\alpha c_{\rm s, mid} H_\mathrm{g}$, with $\alpha$ being direction independent and $c_\mathrm{s}$ describing the midplane sound speed ($c_{s,\rm mid}=\sqrt{k_\mathrm{B} T_{\rm mid}/\mu m_H}$; with the mean molecular weight $\mu=2.3$ and $\mu_\mathrm{H}$ describing the atomic mass of hydrogen). 
The diffusion coefficient for all gas species (which is assumed to be the same in the vertical and radial direction) can be related to $\nu_\mathrm{T}$ using the Schmidt number ($Sc=\nu_\mathrm{T}/D_\mathrm{g}$). We assume a Schmidt number of $1/3$ consistent with other work \citep[e.g.][]{Schneider2021}.
The diffusion coefficient for dust grains (and also ice on these grains) deviates from $D_g$ depending on the grains' Stokes number $\St$:
\begin{align}
    D_\mathrm{d}=\frac{D_\mathrm{g}}{1+ \St^2}.\label{eq:stokes}
\end{align}

The Stokes number is given by 

\begin{align}
    \St = \sqrt{\frac{\pi}{8}}  \frac{s_\bullet \rho_\bullet \Omega}{\rho_\mathrm{g} c_\mathrm{s}}, 
\end{align}

where $s_\bullet$ and $\rho_\bullet$ denote the dust grain size and internal density, respectively. We note that $c_\mathrm{s}$ is the sound speed evaluated using the local grid cell temperature and not the midplane temperature.

To account for advection, we assume a background velocity which is directed towards the star of 

\begin{align}
    \nu_{\mathrm{g}}=\frac{3}{2}\frac{D_\mathrm{g}}{r}.
\end{align}

Using every grid cell position, $\nu_\mathrm{g}$ is split into a vertical $\nu_{z,\mathrm{g}}$ and radial $\nu_{r,\mathrm{g}}$ component. This velocity is used for all gas species that are evolved with Eq.~\ref{eq:advect_diff}. Additionally, we assume that the small dust grains (and ice on small dust grains) are well coupled to the gas and follow the same velocity field. 
The used velocity field does not include strong vertical gradients or e.g. meridional flows \citep[see][]{Ciesla2009}. However, due to the generally shorter timescale of pebble drift and diffusion, the exact shape of the velocity field does not have a strong impact on the results.

The equations are sufficient to model advection and diffusion for water, small dust grains, ice on small dust grains, \ce{OH}, and oxygen by using the initial parameters listed in Table~\ref{tab:model_parameters}.

\subsection{Water desorption and sublimation \label{sec:ice}}

Modelling the region around the water snowline requires a detailed description of the balance between vapour and ice water.
We follow the ice description as detailed by \cite{Krijt2018} and sketched in the top right box of Fig.~\ref{fig:model_sketch}.
In brief, this means that the vapour concentration ($C_\mathrm{H_2O}$) changes according to thermal desorption, photodesorption, and freeze-out:

\begin{align}
    \frac{\partial C_\mathrm{H_2O}}{\partial t} = \frac{3 v_\mathrm{th}}{4 s_\bullet } \frac{\rho_\mathrm{d}}{\rho_\bullet} \bigg[ \overbrace{ \frac{\rho_{\rm sat}}{\rho_\mathrm{g}}  -  \frac{\rho_\mathrm{v}}{\rho_\mathrm{g}} +  \frac{4 m_{\rm H_2O} Y F_{\rm uv}}{v_\mathrm{th} \rho_\mathrm{g}} }^{\mathrm{=-C^*}} \bigg] \, . \label{eq:ice_vapor}
\end{align}

The saturation density is given by:

\begin{align}
    \rho_{\rm sat} = m_{\mathrm{H_2O}} (4 /v_\mathrm{th}) N_\mathrm{s} \times \nu_0 \exp\left\{ - \frac{\mathcal{E}}{ k_\mathrm{B}T} \right\} \,, 
\end{align}
which depends among other things on the binding energy of water ice ($\mathcal{E}/k_\mathrm{B}=5770\,\rm K$), the density of absorption sites ($N_\mathrm{s}=10^{15}\,\rm cm^{-2}$) the thermal velocity $v_\mathrm{th} = \sqrt{8 k_\mathrm{B} T / \pi m_{\rm H_2O} }$, and vibrational frequency $\nu_0=( 2 N_\mathrm{s} \mathcal{E}/\pi^2 m_{\rm H_2O})^{1/2}$.
We use a molecular mass of water ($m_{\rm H_2O}$) of $18$ times the mass of hydrogen.

The photodesorption part of Eq.~\ref{eq:ice_vapor} depends on the local UV field $F_{\rm uv}$ and the photodesorption yield $Y$ which we set to $10^{-2}/\rm photon$ \citep[][]{Westley1995}. The local UV field is determined by $F_{\rm uv}=F_0 e^{-\tau_{\rm uv}}$ with $F_0=\Gamma G_0$. We follow \cite{Krijt2018} by setting $\Gamma=1$ and $G_0=10^{8}\,\rm cm^{-2} s^{-1}$, which accounts for the interstellar radiation field \citep[][]{Habing1968}. The UV opacity includes the vertically integrated opacity of the small dust grains, \ce{H2O}, and \ce{OH} as detailed in Section~\ref{sec:chem}. We note that this description does not account for the stellar UV field the effect of which we discuss in Sect.~\ref{sec:results_chem} and Sect.~\ref{sec:results_alpha}.

The consequent change of water vapour will be matched by an equivalent change of ice concentration:
\begin{align}
\frac{\partial C_\mathrm{H_2O}}{\partial t} = - \frac{\partial C_{\rm ice}}{\partial t} \, .
\end{align}

As shown in Eq.~\ref{eq:ice_vapor}, to solve this equation during every time step we substitute $C^*$ leading to

\begin{align}
    \frac{\partial C^*}{\partial t}=-\frac{3 \nu_{\mathrm{th}} \rho_\mathrm{d}}{4 s_\bullet \rho_\bullet} C^* \, ,
\end{align}
which can be solved as 

\begin{align}
    C^*(t+\Delta t) = C^*(t) \left[1-\exp\left(-\frac{3 \nu_{\rm th} \rho_\mathrm{d}}{4 s_\bullet \rho_\bullet} \Delta t\right)\right]
\end{align}

The water distribution is initialised using the steady-state solution of Eq.~\ref{eq:ice_vapor} (with the additional consideration of ice on pebbles which is discussed in Sect.~\ref{sec:ice_on_peb}). The resulting snowline, defined as the position of equal density between water vapour and ice, is shown in Fig.~\ref{fig:setup}. While the snowline reaches inwards to nearly $0.6\,\rm au$ and $200\,\rm K$ in the midplane, in higher disc regions the snowline is pushed further out (crossing $150\,\rm K$ at $z/r\approx0.2$). This is a consequence of the lower gas densities in higher disc layers (resulting in lower ice fractions due to the density independence of $\rho_{\rm sat}$) and the effect of photodesorption.

\subsection{Chemistry \label{sec:chem}}

Gas-phase water reactions in protoplanetary disc are generally well understood \citep[e.g.][]{vanDishoek2014,vandishoeck2021} and modelled in detail in static thermochemical models \citep[e.g.][]{Woitke2009,Bosman2022}. To allow for the coupling with dynamical processes, we only account for the most important processes that alter the water abundance. These processes are UV-photodissociation (including water self-shielding and OH-shielding) and gas-phase creation of cold water (middle right panel of Fig.~\ref{fig:model_sketch}).

To account for water UV-photodissociation, we define a timescale under which water is destroyed (and transformed into \ce{OH}) by UV photons \citep[$t_{\rm UV,0}=40\,\rm yr$ as introduced for a standard interstellar radiation field by][]{vanDishoek2014}\footnote{This approach is simplifying a timescale that can vary by orders of magnitude \citep[see Fig.~9 in ][]{Vlasblom2025}. However, we show in Sect.~\ref{sec:results_chem} that this approach reasonably approximates more complex results from thermochemical models. We discuss how changes in $t_{\rm UV,0}$ affect the results in Sect.~\ref{sec:results_chem} and Sect.~\ref{sec:results_alpha}}. The applied rate is based based on interstellar UV to stay consistent with previously used chemical survival timescales for water \citep[e.g.][]{Romero-Mirza2024,houge_smuggling_2025}. However, stellar UV can dominate the UV field in the inner disc region leading to photodissociation timescales of $t_{\rm UV,0}\lesssim1\,\rm yr$ \citep[e.g.][]{Vlasblom2025} together with higher photodesorption rates. We discuss this influence when interpreting the results. The photodissociation timescale is only applied in regions where the UV-flux can penetrate. To identify this region we calculate the vertical opacity of the dust, water, and \ce{OH} ($\tau_{\rm UV}=\tau_{\rm UV, Dust}+\tau_{\rm UV, Water}+\tau_{\rm UV,OH}$) at every grid point and update the timescale as $t_{\rm UV}=t_{\rm UV,0}\times \exp(\tau_{\rm UV})$. The timescale is used to reduce the water concentration by $\dot{C}_{\rm Water}=-C_{\rm Water}/t_{\rm UV}$ and calculate the produced \ce{OH}. While all photodissociated water forms  \ce{OH} at high temperatures ($>300\,\rm K$) due to the efficiency of the \ce{OH} to \ce{H2O} forward reaction \citep[][]{Wagner1987}, at low temperature \ce{OH} is largely further processed (by UV radiation into \ce{O}) and therefore does not contribute significantly to the shielding of water \citep{Bethell2009}. We extract the radial \ce{OH} distribution in the disc surface from the standard ProDiMo thermochemical model \citep{Woitke2009} and implement a similar temperature transition in our model. We assume that all photodissociated water above $400\,\rm K$ is transformed into \ce{OH}, below $200\,\rm K$ only $0.5\,\rm \%$ of photodissociated water ends up in \ce{OH} with the remaining majority transformed into \ce{O}, between those regimes the \ce{OH}/\ce{O} fraction produced by water photodissociation follows a power law that matches both end points. We dynamically evolve both \ce{OH} and \ce{O} as independent species. \ce{OH} will deplete into water if sufficiently shielded from UV radiation. Therefore, we introduce a long timescale of $10\times t_{\rm UV,0}$ which translates \ce{OH} back into water if shielded from UV (with a smooth transition between $\tau_{\rm UV}=1$ and $\tau_{\rm UV}=5$).

We calculate the opacity of water and \ce{OH} as $\tau_{\mathrm{UV}}=\sigma  N$, with $N$ denoting the vertical column density from the disc surface to every grid point and $\sigma=\sigma_{\rm water}=\sigma_{\rm OH}=5\times 10^{-18}\,\rm cm^2$ \citep{Bethell2009}.

We set the UV opacity of small dust grains to $10^4\,\rm cm^2/g$, which roughly matches the opacity of $0.5\,\rm \mu m$ grains with the DIANA standard composition \citep{Woitke2016} as calculated using optool \citep{dominik2021}.
This opacity is evaluated at $\lambda_{\rm UV}=0.1\,\rm \mu m$ and integrated vertically downwards using the dust density structure to derive the vertical dust UV opacity. 

\begin{figure}
    \centering
    \includegraphics[width=1.0\linewidth]{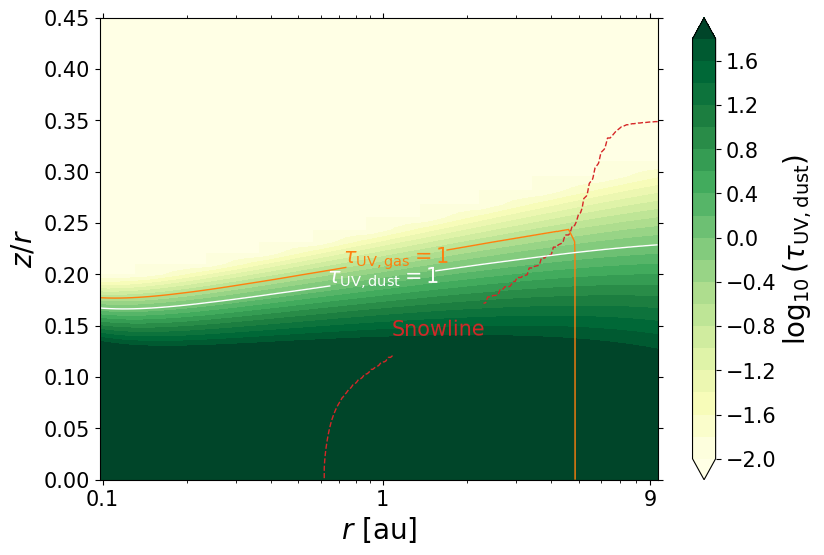}
    \caption{Dust UV opacity of the initial conditions of the C model (top row of Table~\ref{tab:grid}). The snowline is marked in red, with the $\tau_{\rm UV, dust}=1$ and $\tau_{\rm UV, gas}=1$ highlighted in white and orange, respectively.}
    \label{fig:photodis}
\end{figure}

As seen in Fig.~\ref{fig:photodis}, for our initial conditions (Table~\ref{tab:model_parameters}) the dust becomes optically thick between $z/r=0.15$ and $z/r=0.2$. This means that even if water self-shielding is not included or if water is depleted in the upper disc layer, photodissociation will not be effective at disc regions below these heights. The vertical UV opacity of the initialized water density can also be seen in Fig.~\ref{fig:photodis}. The water opacity at the inner part of the disc is higher than the dust opacity leading to effective water self shielding. Beyond the water snowline the water abundance drops quickly leading to an abrupt drop in water opacity. Therefore, the opacity in this region is dominated by the dust particles.

Next to the destruction of water, we also account for the gas-phase formation of water at $T<400\,\rm K$. 
The main reaction pathway  that we model is

\begin{align}
  \ce{O} + \ce{H_3^+} \rightarrow \ce{OH^+} \\
  \ce{OH^+} + \ce{H_2} \rightarrow \ce{H_2O^+} \\
  \ce{H2O^+} + \ce{H_2} \rightarrow \ce{H_3O^+} \\
  \ce{H_3O^+} + \ce{e^-} \rightarrow \ce{H_2O}
  \ .
\end{align}

While most reactions are very fast, the bottleneck is the first reaction including \ce{O} and \ce{H3+}. Therefore, we estimate the amount of water formed as
\begin{align}
    \dot{n}\left(\mathrm{H_2O}\right)=-\dot{n}\left(\mathrm{H}_3^+\right)=n\left(\mathrm{O}\right)n\left(\mathrm{H}_3^+\right)k_{\rm H_3^+,H_2O}\, , \label{reac:form_water}
\end{align} 
with $n$ denoting the abundance and the reaction rate $k_{\rm H_3^+,H_2O}$, which we estimate to be $ 10^{-9}\,\rm cm^3/s $. This means that the abundance of $\mathrm{H}_3^+$ and \ce{O} needs to be known to determine how much water is formed in the gas-phase. We consider one formation and two destruction ways\footnote{We note that we do not update the water abundance according to Reaction~\ref{rec:h2o} since this reaction has only a small effect on the total water abundance.} of \ce{H_3^+}: 

\begin{align}
   \zeta_{\rm CR}+\ce{H2} &\rightarrow \ce{H3+}\\
    \ce{H3+} +\ce{CO} &\rightarrow \ce{HCO+} + \ce{H2} \label{rec:co}\\
    \ce{H3+} +\ce{H2O} &\rightarrow \ce{H3O+} + \ce{H2} \label{rec:h2o}
\end{align}

Assuming a steady state of the abundance of $\mathrm{H}_3^+$, we derive that
\begin{align}
    \frac{\partial n\left(\ce{H3+}\right)}{\partial t}&=0=\zeta_{\rm CR} n\left(\ce{H2}\right) -n\left(\ce{H3+}\right) \left[n\left(\ce{CO}\right)k_1+n\left(\ce{H2O}\right)k_2\right] \\
    n\left(\ce{H3+}\right)&=\frac{\zeta_{\rm CR} n\left(\ce{H2}\right)}{n\left(\ce{CO}\right)k_1+n\left(\ce{H2O}\right)k_2} \label{eq:nh3+_computation}\ .
\end{align}

This equation can be used to calculate the \ce{H3+} abundance using $\zeta_{\rm CR} = 10^{-17} \, \rm s^{-1}$ (in our model), $n\left( \ce{CO} \right)=10^{-4} \times n\left( \mathrm{Gas} \right)$, and $n\left( \ce{H2} \right)=0.5 \times n\left( \mathrm{Gas} \right) $. Thermochemical disc models typically show that \ce{H2} is dominating in lower disc regions with \ce{H} being the dominant hydrogen carrier in the upper layers \citep[e.g.][]{Woitke2022}. We model this transition by removing \ce{H2} from the disc at the upper disc region above the vertical column densities ($N$) of 

\begin{align}
    N_{\rm \ce{H2}-lim} = 5\times 10^{20} \ln\left(43 \frac{G_\mathrm{o}}{n}\right)\,.
\end{align}
with $G_\mathrm{o}$ denoting the strength of the incident FUV field and $n$ being the number density of H-nuclei, following \cite{Tielens2021}. Even though this formula has been derived using constant densities, using $G_\mathrm{o}/n=1$ (resulting in $N_{\rm \ce{H2}-lim}\simeq1.9\times10^{21}\,\rm cm^{-2}$) reproduces the modelled \ce{H}/\ce{H2}-transition in discs well \citep[as seen in Fig.~4 in][]{Woitke2022}. At regions that do not reach $N_{\rm \ce{H2}-lim}$ the \ce{H2} abundance decreases exponentially, crossing the point of $10^{-6}$ at $10^{20}\,\rm cm^{-2}$, which matches the \ce{H2}-abundance slope displayed in Fig.~4 from \cite{Woitke2022}.

The reaction rate $k_1$ (reaction~\ref{rec:co}) and $k_2$ (reaction~\ref{rec:h2o}) for Eq.~\ref{eq:nh3+_computation} are extracted from UMIST \citep{McElroy2013}. Both rates follow the modified Arrhenius equation \citep{Arrhenius1889}. The coefficients are provided in Table~\ref{tab:reactions}.

The resulting abundances of \ce{CO} and \ce{H2} are shown in Fig.~\ref{fig:num_co}  and Fig.~\ref{fig:abund_h2}  in Appendix~\ref{sec:h3plus}. These abundances do not evolve with time, as they are set as constant fractions of the gas density. The same is true for the temperatures-dependent reaction rates for the destruction of \ce{H3+} (Fig.~\ref{fig:water_h3plus_react}), due to the fixed temperature structure that is assumed.

\begin{table*}
    \caption{Reaction rates for \ce{H3+} destruction. The coefficients follow for common notation of the modified Arrhenius equation.}
    \label{tab:reactions}
    \centering
    \begin{tabular}{l|l|l|l|l|l}
    \hline \hline 
         Reaction & $\alpha$ &$\beta$ &$\gamma$ & $T$ limit [K] & Reference \\ \hline
          $\ce{H3+} +\ce{CO} \rightarrow \ce{HCO+} + \ce{H2}$ &$1.36\times 10^{-9}$ &$-0.14$ & $-3.40$ & $10-400$ & \cite{Klippenstein2010}\\ 
          $\ce{H3+} +\ce{H2O} \rightarrow \ce{H3O+} + \ce{H2}$ & $5.90 \times 10^{-9}$ & $-0.50$ & $0.00$& $10-41000$ & \cite{Kim1974,Anicich1975}\\     
    \end{tabular}
\end{table*}

Next to \ce{H3+}, the abundance of \ce{O} is needed to calculate the amount of formed cold water. We assume that the total oxygen abundance which is not locked up in refractory grains is $3.2\times 10^{-4}$ \citep{vandishoeck2021}. From this, we subtract the amount of \ce{CO}, \ce{OH}, and \ce{H2O} (gas and ice) to derive the abundance of \ce{O}. During the simulation two things can change the amount of \ce{O}. First, we assumed that \ce{OH} which is created by photodissociation is transformed into \ce{O} at lower temperature ($T<300\,\rm K$). Second, when water is formed through the reaction of \ce{O} and \ce{H_3^+} this decreases the amount of \ce{O}. These reactions are confined to different parts of the disc (cold photodissociation region and region with high \ce{H_3^+} abundance). Therefore, the advection-diffusion equation Eq.~\ref{eq:advect_diff} is solved for \ce{O} as well.

\subsection{Pebbles \label{sec:pebbles}}

Next to small dust grains and different gas species, we evolve a pebble population that connects via growth and fragmentation to the dust and transports water ice from the outer to the inner disc through radial drift.

We assume that all pebbles are settled and effectively confined to the midplane due to the short settling timescale. Therefore, instead of modelling the 2D distribution, the surface densities are modelled. The pebble scale height ($H_\mathrm{p}$) can be calculated based on the gas scale height ($H_\mathrm{g}$) according to \cite{Birnstiel2024} as

\begin{align}
    H_\mathrm{p}=H_\mathrm{g} \sqrt{\frac{\alpha}{\St+\alpha}}\, . \label{eq:peb_scale}
\end{align}

The surface density changes between dust grains and pebbles are vertically distributed using Eq.~\ref{eq:peb_scale}. 

Pebble densities are governed by fragmentation, coagulation, radial drift, diffusion, and parametrised influx of pebbles from the outer disc

\begin{equation}
\begin{split}
    \frac{\partial}{\partial t} \Sigma_{\mathrm{p}} = &\overbrace{\frac{\Omega\Sigma_{\mathrm{d}} \Sigma_{\mathrm{d}}}{\Sigma_\mathrm{g}}}^{\rm Coagulation}-\overbrace{f_{\rm col}\frac{\Sigma_\mathrm{p}}{t_{\rm col}}}^{\rm Fragmentation} -\overbrace{\frac{1}{r}\frac{\partial}{\partial r}\left[r \Sigma_{\mathrm{p}} \nu_{\rm Drift}\right]}^{\rm Drift}  \\
    &+\underbrace{\frac{1}{r}\frac{\partial}{\partial r}\left[r D_\mathrm{p} \Sigma_{\mathrm{g}}\frac{\partial}{\partial r}\left(\frac{\Sigma_\mathrm{p}}{\Sigma_\mathrm{g}}\right) \right]}_{\rm Diffusion} +\underbrace{\dot{\Sigma}_{\rm in}}_{\rm Influx} \, .
   \label{eq:peb_evo}
\end{split}
\end{equation}
with $\Sigma_{\mathrm{p}}$ being the pebble surface density, $\nu_{\rm Drift}$ being the pebble drift velocity, $t_{\rm col}$ being the collisional time scale of pebbles, $D_\mathrm{p}$ denoting the diffusion coefficient for pebbles (Eq.~\ref{eq:stokes}) and the tunable parameter $f_{\rm col}$ adjusts the collisional timescale of pebble to account for some of the simplifying assumption of our 1D approach (details below). $\dot{\Sigma}_{\rm in}$ is the parametrised influx of pebbles from the outer disc and only affects the outermost disc cell. We set the constant pebble influx ($\sim 2.4\times 10^{-4}\,\rm M_{\oplus}/yr$) to replenish the material which is drifting inwards from the outermost cell at the beginning of the simulation and keep it constant over time.

We calculate the drift velocity as 

\begin{align}
    \nu_{\rm Drift}=-2\eta r \Omega \frac{\St}{1 + \St^2}\,,\label{eq:drift}
\end{align}
with $\eta$ denoting the dimensionless pressure gradient evaluated in the disc midplane, which is given by

\begin{align}
    \eta= \frac{1}{2} \left(\frac{c_\mathrm{s}}{r\Omega}\right)^2\frac{\partial \ln{\rho_\mathrm{g}}}{\partial \ln{r}} \, . \label{eq:pressure_grad}
\end{align}

All coagulated dust grains and fragmented pebbles that transition between the pebble surface density and dust grain density are subtracted/added to the other population to conserve mass.

Since dust in the inner disc is generally fragmentation limited in size \citep{Birnstiel2012}, the representative Stokes number of the large dust grain population can be written as:

\begin{align}
    \St_{\rm frag}=\frac{f_{\rm frag}}{3}\frac{\nu_{f}^2}{\alpha c_\mathrm{s}^2}\,, \label{eq:stokes_peb}
\end{align}
with $f_{\rm frag}\approx0.37$ \citep{Birnstiel2012}, and $\nu_{\mathrm{f}}$ denoting the fragmentation velocity, which we assume to be $5\,\rm m \,s^{-1}$. 
This is related to the particles size via

\begin{align}
    a_{\mathrm{p}}= \frac{2\St\, \Sigma_\mathrm{g}}{\rho_\mathrm{p} \pi}\,,
\end{align}
with $\rho_\mathrm{p}$ denoting the internal pebble particle density which is set to the same value as the internal dust grain density $\rho_\bullet$ (assumed to be $1.6\,\rm g cm^{-3}$).

The collisional time scale depends on the midplane pebble number density ($n_{\mathrm{p}}$), the collisional cross section ($\sigma_{\rm col}$), and the relative particle velocity ($\nu_{\rm rel}$):
\begin{align}
    t_{\rm col} = \frac{1}{n_\mathrm{p} \sigma_{\rm col} \nu_{\rm rel}}\,.
\end{align}

While the local number density can be calculated assuming a Gaussian vertical distribution of particles with a scale height $H_{\mathrm{p}}$ according to Eq.~\ref{eq:vert_distr}, the cross section depends on the particle radius ($a_{\rm frag}$) as $\sigma_\mathrm{col}=4\pi a_\mathrm{p}^2$. Additionally, we assume that the representative relative velocity is given by the fragmentation velocity $\nu_{\rm frag}$. As shown in Eq.~\ref{eq:peb_evo}, the pebble collisional timescale is adjusted using the tunable $f_{\rm col} $ parameter. This is done to match pebble/small dust ratios as predicted by other models. \cite{Birnstiel2012} find a typical mass ratio between pebbles and small dust grains of $0.75$ in the fragmentation limited case. In our model, we set $\Sigma_\mathrm{d}/\Sigma_p=1/3$, by introducing a radius dependent tunable parameter which is calculated by assuming a steady state between coagulation and fragmentation in Eq.~\ref{eq:peb_evo}:
\begin{align}
        f_{\rm col}=\underbrace{\left(\frac{\Sigma_\mathrm{d}}{\Sigma_\mathrm{p}}\right)^2}_{\left(1/3\right)^2}\frac{\Omega \sqrt{2\pi}H_{\mathrm{p}}m_\mathrm{p}}{\Sigma_\mathrm{g} \sigma_{\rm col}\nu_{\rm frag}}\,. \label{eq:fcol}
\end{align}

While $f_{\rm col}$ is radius dependent, for our model $f_{\rm col}(1\,\rm au)$ is approximately $0.04$, meaning that the collisional timescale of pebbles is $25$ times longer than estimated using midplane densities and fragmentation velocities as relative velocities. 

\subsubsection{Ice on pebbles\label{sec:ice_on_peb}}

Next to the pebble surface density, we also evolve the (surface) density of ice on pebbles. The ice on pebbles surface density is evolved using the drift, diffusion, and influx as seen in Eq.~\ref{eq:peb_evo}. The parametrised influx of pebbles ($\dot{\Sigma}_{\mathrm{p}}$) is assumed to carry a fraction of water ice which is equivalent to the simulation's initial water to solid ratio.
Additionally, transforming dust into pebbles and vice versa changes the ice concentrations of ice on dust grains and ice on pebbles. The amount of transferred ice is calculated using the surface density of coagulated dust grains and fragmented pebble and the ice to grain ratio of the transformed material ($\rho_{\rm ice}/\rho_{\rm dust}$ in the local cells for dust or $\Sigma_{\rm ice}/\Sigma_{\rm peb}$ for pebbles). This means that even in a steady state between dust coagulation and pebble fragmentation, the ice/pebble ratio can change if the ice/pebble and ice/dust ratio are not identical. To translate the ice surface density into local densities, we assume that the ices on pebbles are vertically distributed by a Gaussian with the pebble scale height and that all pebble/small-grain interactions happen within this distribution.

Ice on pebbles interacts also with the water vapour through freeze out and desorption (equivalent to Eq.~\ref{eq:ice_vapor}) with all parameters relating to the dust ($\rho_{\rm d}$, $s_\bullet$, and $\rho_\bullet$) being replaced by their pebble counterparts\footnote{For simplicity we assume the same internal density for pebbles and dust grains.}. Due to the larger relative surface area of all dust grains compared to pebbles, we decide to solve Eq.~\ref{eq:ice_vapor} for dust first and only use any remaining water vapour for freeze-out and desorption of ice on pebbles.

\subsection{Numerical implementation\label{sec:numerics}}

\subsubsection{Spatial grid setup\label{sec:num_model_setup}}

We use a logarithmically spaced grid in $r$ and $z$ for which $r_{i+1}/r_{i}=1.05$ and $z_{i+1}/z_{i}=1.05$. We model the radial region between $0.1-10\,\rm au$ which contains the water snowline for all our temperature structures. While water emission is likely to originate at closer-in regions as well, we account for this during the spectrum generation (Sect.~\ref{sec:sim_obs}). The disc is modelled up to $6$ gas scale heights. We tested including higher disc regions but none of the results in this paper changed significantly. This is because (1) the water emission from higher disc layers is unlikely to significantly contribute to the total emission due to the rapidly vertically decreasing densities and (2) water is rapidly photodissociated in the optically thin disc surface leading to naturally low abundances that are not replenished by chemistry above the H-H$_2$ transition.
The lowest vertical grid point is centred at $0.005\,\rm au$ and has an extended lower limit at the disc's midplane. 

\subsubsection{Boundary conditions \label{sec:bound}}

The boundary conditions are reflective in the vertical ($\partial C_i/\partial z=0$) and radial direction ($\partial C_i/\partial r=0$) for all modelled species (including pebbles). Consequently, diffusion does not change the total mass in the modelled region. However, we allow for advection to remove material through the inner boundary of the simulation and add material at the outer edge as well. This transparency of the radial boundaries is enforced by refilling the boundary cells with the material they exchange through advection with their adjacent cells.

\subsubsection{Time step \label{sec:num_time-step}}

The model evolution is done using a finite-difference method based on an evolving time step $\Delta t$. This time step is adjusted during the simulation based on $\Delta t_{\rm new}$ on the maximum expected concentration change of all evolved species in every grid cell and the maximum surface density change. We only allow for $10\,\%$ of material from every cell to change location per time step by introducing a factor $f_\mathrm{c}=0.1$ that decreases the time step. After initialising $\Delta t$ as $f_\mathrm{c}\times \Delta t_{\rm new}$, for every time step with $f_\mathrm{c} \times\Delta t_{\rm new} >  \Delta t$, the simulation is evolved for $\Delta t$, which is subsequently increased by $5\, \%$. If  $f_\mathrm{c} \times \Delta t_{\rm new} < \Delta t$, the simulation evolves for $f_\mathrm{c} \times \Delta t_{\rm new}$, with $\Delta t$ being decreased by $5\,\%$ for the next time step. This algorithm results in a smooth change of time steps with drastic decrease if the concentration/surface density changes are too high for the cell's concentration/surface densities. In practise this results in typical time steps on the order of $\sim 10\,\rm days$ if all evolutionary processes are included, which can increase to values up to a few years in the chemistry-only case without any dynamics\footnote{If only chemistry is considered the time step is in most cases given by $f_c$ times the photodissociation timescale introduced in Section~\ref{sec:chem}, which acts as a natural maximum possible timestep.}. All simulations run for $1\,\rm Myr$. We deem this time sufficient to examine all the processes with their shorter timescales.

\begin{table*}
    \centering
    \caption{Grid of models. The individual columns denote which processes are included in the model.\label{tab:grid}}
    \begin{tabular}{lccccc}
    Model                   & Chemistry & \multirow{2}{1.4cm}{Diffusion/ Advection} &  \multirow{2}{1.8cm}{Coagulation/ fragmentation} & \multirow{2}{1.85cm}{Pebble drift/ Pebble diffusion} & Results \\ 
    & & & & &  \\ \hline \hline
    
    \textbf{Main models}    & & & & &  \\
    C(hemistry)          & \checkmark &            &                     &           &  Sect.~\ref{sec:results_chem} \\
    C+D(iffusion \& advection) & \checkmark & \checkmark &                     &           & Sect.~\ref{sec:chem_dyna}  \\

    C+D+Co(agulation \& fragmentation)           & \checkmark & \checkmark &  \checkmark & &  Sect.~\ref{sec:results_no_drift} \\
    C+D+Co+Dr(ift \& pebble diffusion)          & \checkmark & \checkmark & \checkmark & \checkmark &  Sect.~\ref{sec:results_drift} \\ \hline 
    \textbf{Supporting models}    & & & & &  \\
    C+D with $\alpha=10^{-4}$ & \checkmark & \checkmark &                 &           & Sect.~\ref{sec:chem_dyna}  \\
    C+D+Co+Dr with $2\times v_{\mathrm{Drift}}$ & \checkmark & \checkmark & \checkmark                  & \checkmark          & Sect.~\ref{sec:results_drift}  \\
    C+D+Co+Dr with $v_{\mathrm{Drift}}/2$ & \checkmark & \checkmark & \checkmark                  & \checkmark          & Sect.~\ref{sec:results_drift}  \\
    C+D+Co+Dr with $v_{\mathrm{frag}}=2.5-5\,\rm m/s$            & \checkmark & \checkmark & \checkmark & \checkmark       &  Sect.~\ref{sec:results_drift} \\
    C+D+Dr                & \checkmark & \checkmark &  & \checkmark       &  Sect.~\ref{sec:results_drift} \\  \hline             
    \end{tabular}
\end{table*}

\begin{figure}
    \centering
    \includegraphics[width=0.9\linewidth]{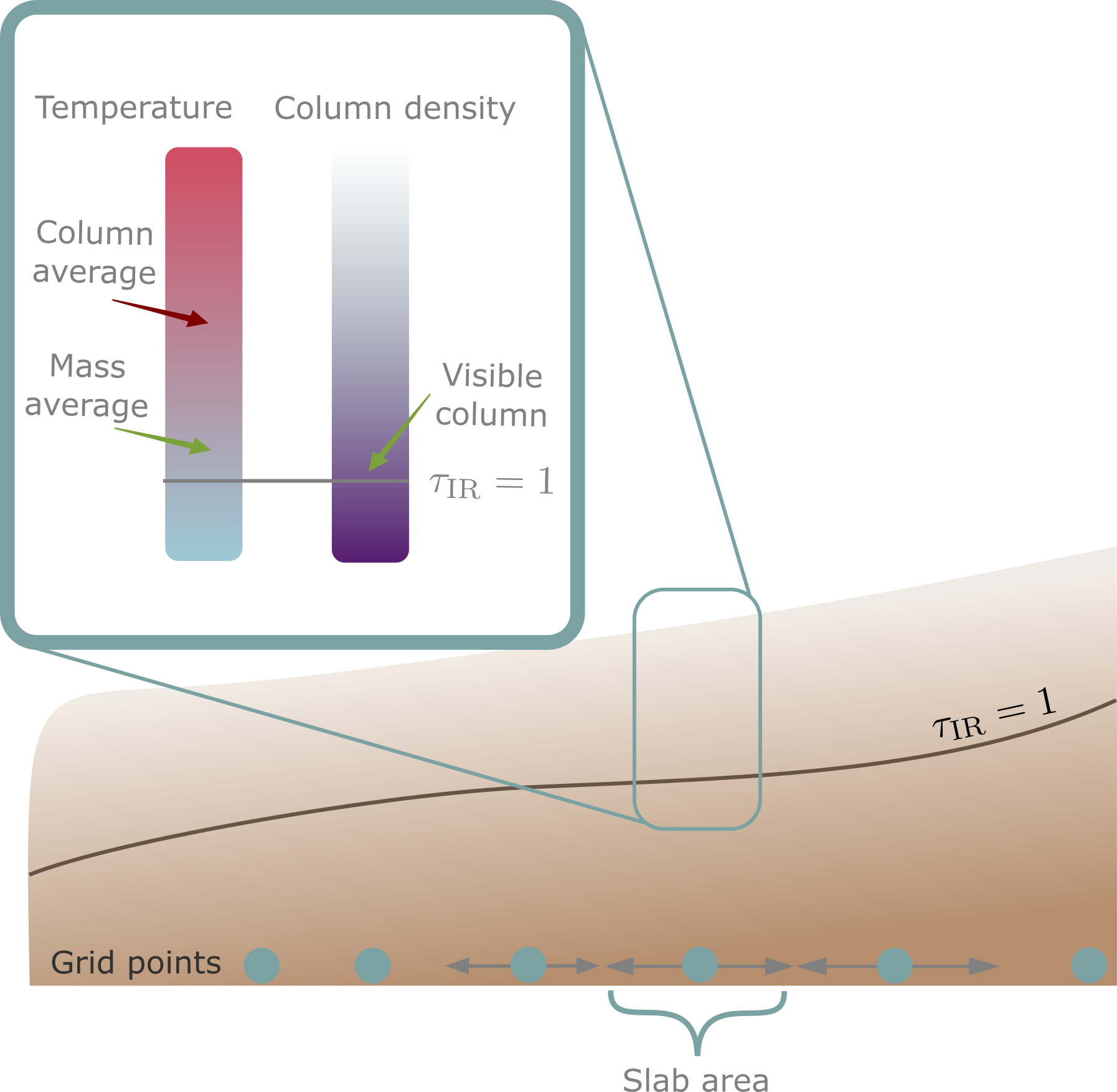}
    \caption{Sketch of the slab parameter extraction from the 2D disc structure. The emission is modelled using a radial series of slab models, each representing one grid point using the slab area, column density to $\tau_{\mathrm{IR}}=1$ in the mid-IR and the mass averaged temperature in that visible column.}
    \label{fig:slab_extraction}
\end{figure}

\subsection{Simulated observations \label{sec:sim_obs}}

Based on the water density structure at any given time step we simulate mid-IR water spectra as seen by JWST/MIRI. To do so, we use the Dust Continuum Kit with Line Emission from Gas \citep[DuCKLinG;][]{Kaeufer2024}. This tool uses a large temperature and column density grid of Local thermodynamic equilibrium (LTE) slab models \citep{Arabhavi2024} based on data from the HITRAN 2020 database \citep{Gordon2022}. The slab models include line overlap, are convolved with an approximate MIRI resolution ($R=3000$), and rebinned to a typical MIRI wavelength grid. We use DuCKLinG to efficiently interpolate in this grid and calculate the flux for any given combination of temperature, column density, and emitting area.

While slab models use many simplifications, it has been shown that when using multiple slab models fluxes from complex 2D density structures can be sufficiently approximated \citep{Kaeufer2024,Vlasblom2025}. Therefore, we use one slab model for every radial grid point in our 2D density structure (Fig.~\ref{fig:slab_extraction}). The emitting area corresponds directly to the area of the annulus represented by the radial grid cell position. The column density to which JWST/MIRI is sensitive includes all water vapour above the $\tau_{\rm IR}=1$ line. This line evolves with changes of the dust density and is evaluated using a dust opacity of $5\times10^3\,\rm cm^{2}/g$ (roughly matching the opacity calculated with optool at $10\,\rm \mu m$ as described in Sect.~\ref{sec:chem}). From this observable water column, the mass averaged temperature is calculated to complete the set of three parameters needed by a 0D slab model to simulate the water flux. All these 0D slab model fluxes are then added to derive the total mid-IR water spectrum.

Instead of analysing the full water spectrum, a small selection of well understood unblended water lines and their ratios can also be used to analyse the water emission coming from a disc \citep{Banzatti2025}. Therefore, we extract the same quantities from our simulated spectra. For doing so, the fluxes of the lines introduced by \cite{Banzatti2025} are integrated and their ratios calculated (more details in Sect.~\ref{sec:results_spec}).

\subsection{Grid of models\label{sec:grid}}

While a full parameter exploration is beyond the scope of this paper, we set up a small grid of four main models to explore the effect of different included processes on the 2D water density structure and simulated observations.
These models are listed in Table~\ref{tab:grid} with their included processes. All parameters that are the same for all models are listed in the parameter table (Table~\ref{tab:model_parameters}). This means that these models are initialised with the same underlying gas, dust, water, \ce{CO}, \ce{H2}, oxygen, and pebble density structure (even if pebbles are not used during the model's evolution). While model~C+D+Co+Dr includes all processes, the model without pebble drift and diffusion (model~C+D+Co) can be interpreted as a model with negligible pebble flux. Similarly, model~C+D+Co simplifies into model~C+D if the Stokes number of large grains is smaller than $\alpha$. In that case, no settled large dust grains exist, leaving only diffusion, advection, and chemistry to determine the 2D water density structure. Lastly, if the turbulence in a disc is low, model~C+D can be approximated by model~C. In this case all dynamical timescales are longer than the chemical timescales. 

We evolved five further models in addition to the set of four main models. These evaluate the influence of the $\alpha$ parameter (Sect.~\ref{sec:chem_dyna}), drift velocity (Sect.~\ref{sec:results_drift}), changing fragmentation velocity (Sect.~\ref{sec:results_drift}) and the feedback loop between coagulation/fragmentation and pebble drift (Sect.~\ref{sec:results_drift}).

\section{The effect of different processes on water concentrations}
\label{sec:results_density}

In this section, we highlight the influence of different processes on the 2D water density and 1D surface density structure. We start with the static chemical model (model~C), add diffusion and advection (model~C+D), add coagulation and fragmentation (model~C+D+Co), and finally pebble drift and diffusion (model~C+D+Co+Dr). All models are listed in Table~\ref{tab:grid}.

\subsection{Chemical processing of water (model~C)\label{sec:results_chem}}

\begin{figure*}
    \centering
    \includegraphics[width=0.95\linewidth]{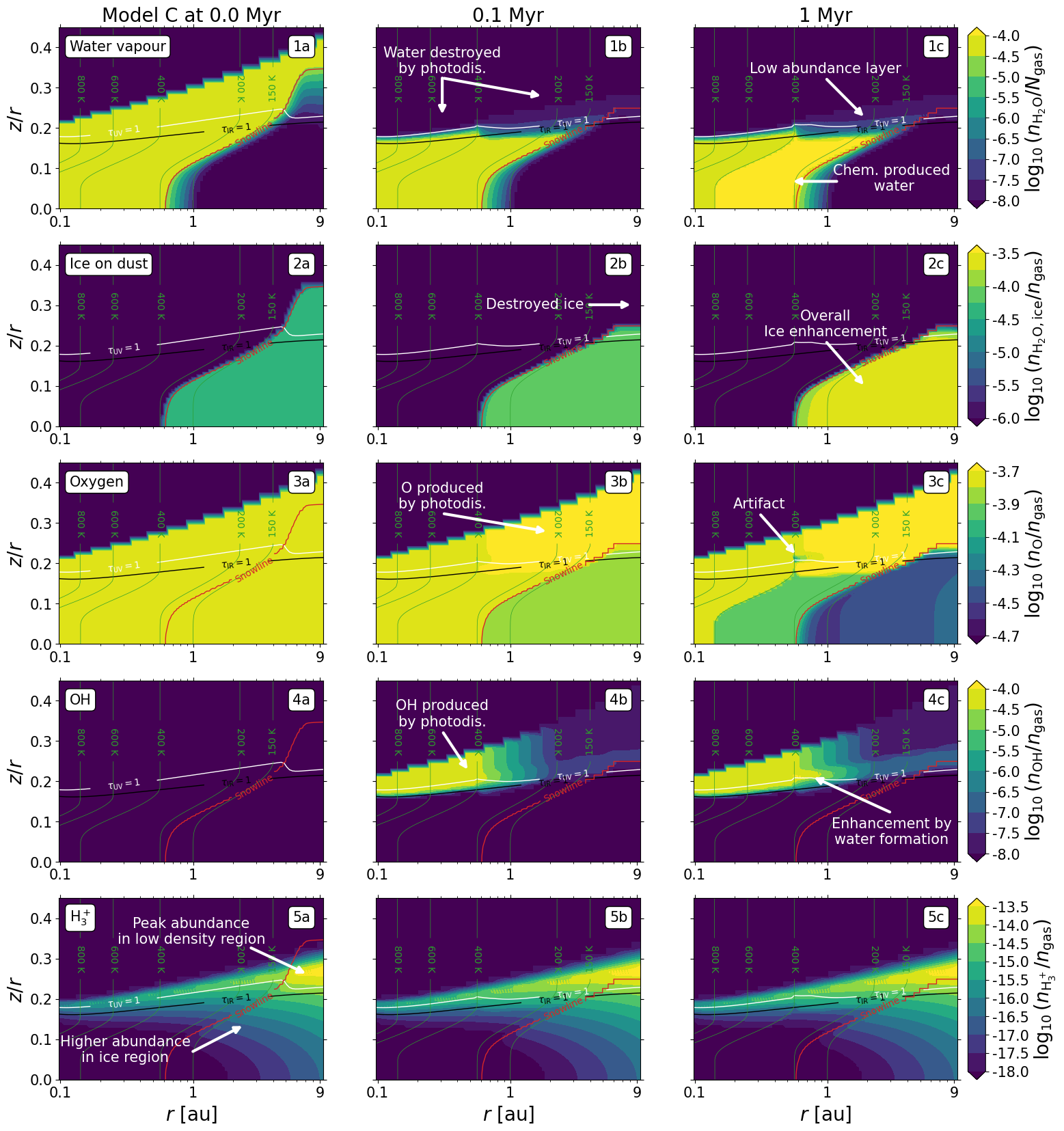}
    \caption{The chemical evolution of water in model~C (row one of Table~\ref{tab:grid}) . Every column depicts a different time during the evolution (initial conditions, $0.1\,\rm Myr$, and $1\,\rm Myr$) of the water abundance (top row), water ice abundance (second row), oxygen abundance (third row), \ce{OH} abundance (fourth row), and \ce{H3+} abundance (last row). The snowline, temperature contours and opacity lines are shown for reference.}
    \label{fig:chem_2d}
\end{figure*}

The evolution by gas phase water formation and photodissociation of the static chemistry model without dynamics (model~C) is depicted in Fig.~\ref{fig:chem_2d}. The left panels show the initial conditions (panels~1a-5a). \ce{H2O} (vapour (1a) plus ice (2a)) and oxygen (panel~3a) are initialized as a constant abundance throughout the model (see Sect.~\ref{sec:chem}). The snowline, which is defined as equal water vapour and ice densities \citep[for a discussion on different definitions see][]{Calahan2026}, separates the two water phases. While water ice does practically not exist inside the water snowline (panel~2a) a low fraction of water vapour exists beyond the snowline (panel~1a, equivalent to $\rho_{\rm sat}$ in Eq.~\ref{eq:ice_vapor}). No \ce{OH} exists at the beginning of our simulation (panel~4a) as it is exclusively formed via water photodissociation in our simulation. The initial \ce{H3+} abundance (panel~5a) decreases in water-rich regions due to \ce{H3+} destruction via water (Fig.~\ref{fig:water_h3plus_react}). This means that at the initial conditions \ce{H3+} is more abundant at higher disc layers especially outside the water snowline. Nevertheless, the peak abundance of \ce{H3+} is very low at about $10^{-13}$. We show in Appendix~\ref{sec:h3plus} that the \ce{H3+} abundance increase is largely driven by the decrease in total gas density and consistent with previously found behaviour of \ce{H3+} in the  interstellar medium \citep{Oka2013}. Therefore, gas-phase water formation via \ce{H3+} is more important in less massive discs. 

When photodissociation and gas-phase water formation are given time ($>0.1\,\rm Myr$) to influence the molecular abundances (column~b and c of Fig.~\ref{fig:chem_2d}) stark changes in abundances can be noted. Photodissociation reacts on very short timescales and therefore reduces the amount of water vapour (panel~1b) in the UV exposed layer early on. In the inner region, this photodissociated water transforms into \ce{OH} (panel~4b) which provides its own UV opacity leading to a constant overall UV opacity in the inner region over time (e.g. panel~4b and 4c). Outside of the $T=400\,\rm K$ contour, an increasingly large fraction of water is photodissociated into \ce{O} instead of \ce{OH} (panel~3b), which does not contribute to the UV opacity leading to a lowering of the $\tau_{\rm UV}=1$-line. Therefore, in these regions the dust opacity (compare e.g. panel~3b of Fig.~\ref{fig:chem_2d} to Fig.~\ref{fig:photodis}) dominates. 

Additionally, lowering the amount of water in the disc surface leads to an increase in the \ce{H3+} abundance (panel~5b and 5c). Both the increase in \ce{O} and \ce{H3+} result in higher production rates of water through gas-phase chemistry (Reaction~\ref{reac:form_water}). Below the $\tau_{\rm UV}=1$ line photodissociation plays a minimal role, leading to an effective transformation of \ce{O} into \ce{H2O} (panels~3b and 3c). This reaction is faster outside the snowline due to lack of gas-phase water which would reduce the \ce{H_3^+} abundance. This can be seen in panel~3c that shows the strongest \ce{O} depletion beyond the water snowline. Additionally, the oxygen abundance decreases around the $\tau_{\rm UV}=1$ line close to $400\,\rm K$. This is due to the constant water production in this region, which is quickly photodissociated (mainly into \ce{OH} instead of \ce{O}). We note that the upper boundary of this region consists of the H/\ce{H2}-transition. This specific artifact is attributed to the limitations imposed by our semi-analytical chemistry network (e.g. the temperature limit for cold water formation) and is not expected to show up in models using full chemical networks. However, the water abundance in this region, which is the main subject of this study, does not follow this trend which leaves us to accept the oxygen behaviour as a necessary and inconsequential effect of the model's simplification. 

The chemically produced \ce{H2O} vapour can be seen below the $\tau_{\rm UV}=1$ line at temperatures below $400\,\rm K$ (panel~1c). Additionally, panel~2c shows that produced water vapour outside the snowline freezes out quickly, resulting in higher ice abundances compared to the initial conditions. 
While all water above the $\tau_{\rm UV}=1$ line at temperatures larger than $400\,\rm K$ is photodissociated, the water production is acting against this destruction at colder regions. Therefore, \ce{H2O} exists at abundances of $\sim 10^{-6}$ above the snow surface (panel~1b and 1c). This coincides with the strong reduction in ice abundance above the $\tau_{\rm UV}=1$ line (panel~2b). In this region water vapour is photodissociated, resulting in a redistribution of ice and vapour in that region, which results in a net ice loss. 

We note that the water abundance above the snow surface is very sensitive to the photodissociation rate. In case of higher UV fields, the balance between chemical water formation and photodissociation results in lower water abundances in the disc surface \citep[e.g.][]{Vlasblom2025,Calahan2026}. Similarly, increases in photodesorption rates reduce the ice abundance in the surface. 

To summarise, in the absence of any dynamics, the chemical network leads to (1) a photodissociated surface layer in which water is depleted, (2)
a layer of \ce{OH} that, together with water self-shielding, and dust shielding protects the lower water reservoir from photodissociation, (3) an increase of water vapour and ice in regions outside $400\,\rm K$ through cold water formation, and (4) a tail of $\sim10^{-6}$ water abundance above the snow surface produced via constant photodissociation and water formation.

These seen water reservoirs generally align with the abundances retrieved from complex thermochemical models \citep[e.g.][]{vandishoeck2013,Woitke2024,Vlasblom2025}. For example, Fig.~D.1 in \cite{Arabhavi2024} depicts a similar low abundance water vapour reservoir above the midplane water reservoir for a grid of $25$ thermochemical models. This means that even though our simple chemical approach does not capture all details of thermochemical models, it reproduces the key pathways for water formation and destruction while allowing us to add various dynamical processes.

\subsection{The interplay between chemistry and diffusion (model~C+D)\label{sec:chem_dyna}}

\begin{figure*}
    \centering
    \includegraphics[width=0.95\linewidth]{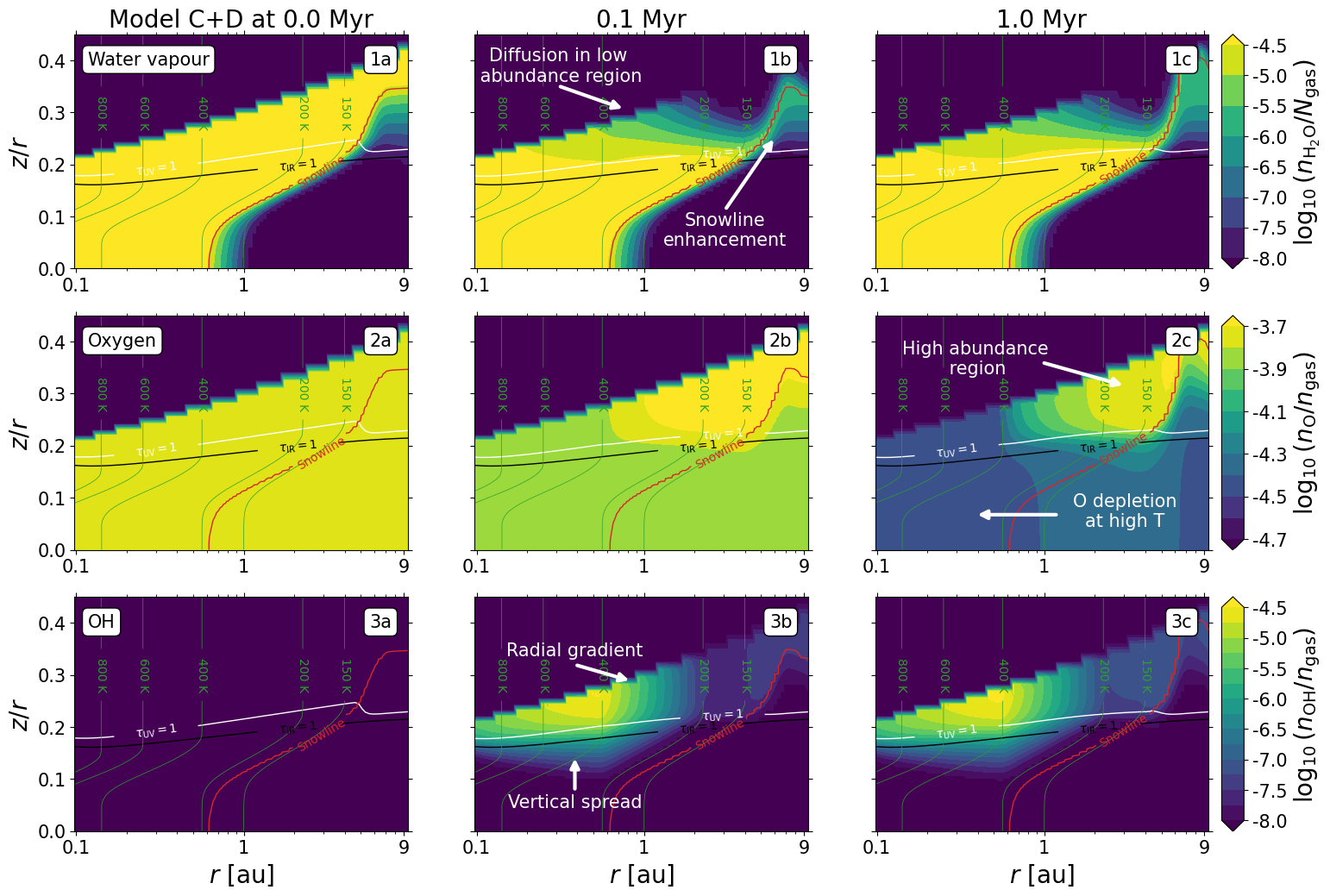}
    \caption{The chemical and dynamical evolution of water in model~C+D (second row of Table~\ref{tab:grid}). Every column depicts a different time during the evolution (initial conditions, $0.1\,\rm Myr$, and $1\,\rm Myr$) of the water abundance (top row), oxygen abundance (second row), and \ce{OH} abundance (last row). The snowline, temperature contours and opacity lines are shown for reference.}
    \label{fig:chem_and_dyna_2d}
\end{figure*}

Next, we are examining how adding diffusion and advection to the chemical processing (model~C+D; second row in Table~\ref{tab:grid}) changes the 2D abundances (Fig.~\ref{fig:chem_and_dyna_2d}). Generally, diffusion acts to reduce concentration gradients leading to increases in regions of low concentration, and vice versa in regions of high concentration if the diffusion timescale is smaller than the chemical reprocessing \citep{Semenov2011}. This can be seen most dramatically for oxygen (panel~2a-2c of Fig.~\ref{fig:chem_and_dyna_2d}). While the regions with temperatures too high for gas-phase cold water formation stayed unaffected when including chemistry only (compare to Fig.~\ref{fig:chem_2d}), the oxygen abundances in these regions change due to the inclusion of diffusion (panel~2c). The abundances decrease steadily over time because of ongoing water formation in the cold regions. The only region that shows a strong \ce{O} abundance gradient after $1\,\rm Myr$ (panel~2c) sits above the $\tau_{\rm UV}=1$ line where (1) water is continuously destroyed leading to an increase of oxygen and (2) at least partly \ce{H2} does not exist stopping the oxygen destruction via cold water formation. However, this region which is still shaped by e.g. vertical diffusion (panel~2b and 2c), but retains a strong concentration gradient.

While the peak \ce{OH} abundances in the chemistry only model is confined to the warm regions of the disc surface, gas dynamics redistribute \ce{OH} leading to a vertical and radial spread (panel~3b). The vertical extent that \ce{OH} exists below the $\tau_{\rm UV}=1$ line depends on the ratio of the vertical diffusion timescale and the assumed timescale for the \ce{OH} to \ce{H2O} reaction. We note that the assumed timescale decreases the \ce{OH} abundance to $<10^{-8}$ at $z/r=0.1$. Therefore, the back reaction fulfils the attempted purpose of stopping a large \ce{OH} build-up near the midplane. Shortening the timescale further would only marginally change the vertical location of \ce{OH}-produced water, which quickly diffuses vertically. The radial spread means that the UV opacity increases outside of the $T=400\,\rm K$ contour which leads to a decrease in water photodissociation.

The water abundance in the disc surface is significantly increased by the inclusion of dynamics compared to the chemistry only case. As seen in panel~1b, diffusion acts against the constant water destruction which increases the abundance in these regions. This is consistent with several studies showing that adding a vertical mixing term to (thermo)chemical models significantly increases the water abundance in the disc surface \citep{Furuya2013,Woitke2022}. Also, water is efficiently retained close to the snowline in the surface layer (panel~1b). This water reservoir originates from sublimating water ice. While in regions further from the disc surface ice diffusion at the snowline is matched by an equal water vapour diffusion into icy regions, the removal of water vapour from the disc surface means that sublimating ice now efficiently supplies the gas-phase.

In summary, adding diffusion to a chemical model (1) reduces abundance gradients for all dynamically evolving species and (2) leads to higher water abundances in the photodissociation region (especially close to the water snowline).

\begin{figure}
    \centering
    \includegraphics[width=0.95\linewidth]{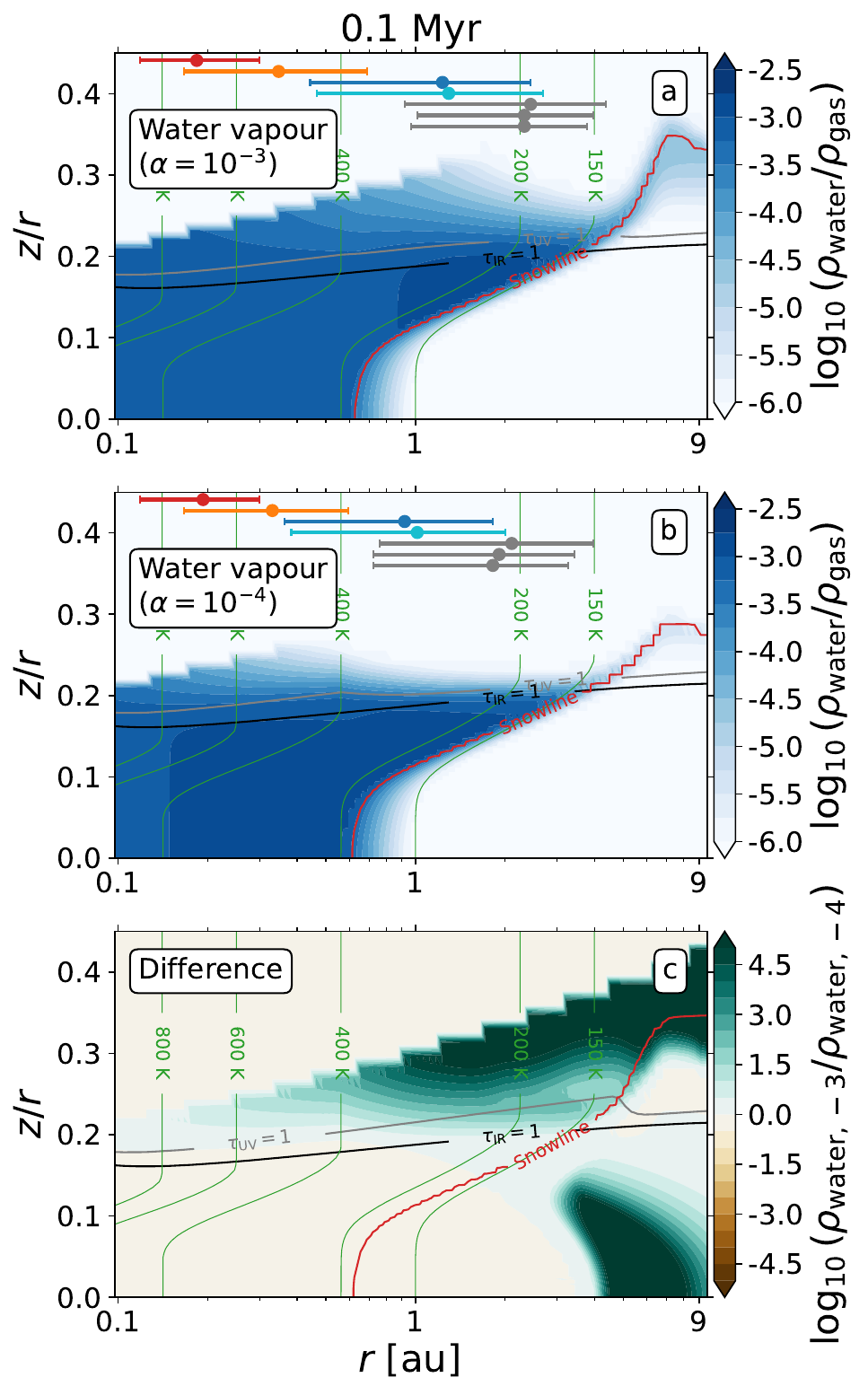}
    \caption{The water vapour concentration in model~C+D with $\alpha=10^{-3}$ (panel~a) and $\alpha=10^{-4}$ (model~C+D with $\alpha=10^{-4}$, panel~b) at $0.1\,\rm Myr$. The difference between those models is shown in panel~c. The overplotted contours in all difference panels are taken from the initial conditions. The horizontal lines in panels~a and b show the emitting regions of the diagnostic water lines (more details in Sect.~\ref{sec:results_spec}) and far-IR lines (more details in Sect.~\ref{sec:possible_solutions}).}
    \label{fig:pebble_compare_alpha}
\end{figure}

\subsubsection{The impact of the $\alpha$-parameter}
\label{sec:results_alpha}

The extent and strength of the water enhancement in the surface depends on the ratio between the (local) dynamics (diffusion and advection) and the chemistry timescale, which \cite{Semenov2011} define as the Damköhler number $\rm Da$. If $\mathrm{Da}\gg1$ chemical evolution is fast and dynamical effects are negligible. On the other hand, $\mathrm{Da}\lesssim1$ means that the concentration is sensitive to dynamical processes. The dynamical timescale and therefore $\mathrm{Da}$ is sensitive to the chemical timescales and the disc turbulence controlled by $\alpha$. Therefore, different photodissociation rates and $\alpha$ values lead to strong changes of the water concentration in the disc surface \citep[e.g.][]{Furuya2013}.

In the following, we test the impact of different $\alpha$ values with a full exploration of different chemical timescales beyond the scope of this work.
Fig.~\ref{fig:pebble_compare_alpha} compares the fiducial case of model~C+D with $\alpha=10^{-3}$ with another model~C+D with $\alpha=10^{-4}$ (first supporting model in Table~\ref{tab:grid}). We compare these models at $0.1\,\rm Myr$ since Fig.~\ref{fig:chem_and_dyna_2d} shows that the balance between diffusion and photodissociation is established at that time already. A lowering of $\alpha$ reduces the strength of diffusion, which in turn results in lower water abundances in the surface layer. While the warm disc surface ($T>400\,\rm K$) exhibits concentration reductions within factors of $~\sim10$, the cold surface shows reductions by more than a factor of $\sim10^5$. This is generally consistent with \cite{Woitke2022}, who showed that including vertical diffusion with $\alpha=10^{-3}$ can enhance the water concentration in the disc surface at a radius of $10\,\rm au$ by more than $6\,\rm dex$. Lastly, we note that the water concentration close to the midplane (below about $ \tau_{\rm UV}=1$) is higher (within a factor $<10$) for model~C+D with $\alpha=10^{-4}$. This is because this water efficiently diffuses into the low concentration regions in the $\alpha=10^{-3}$ case while being retained at lower $\alpha$.

The Damköhler number describes the balance between chemical and dynamical evolution. Therefore, changes to the chemical timescale can lead to similar concentration changes as discussed in the case of changing $\alpha$ values. Increasing the photodissociation rate (e.g. due to a stronger UV field) increases $\rm Da$ which minimises the impact of vertical diffusion to increase water concentration in the disc surface.

\subsection{Effect of non-drifting pebbles on water distributions (model~C+D+Co)\label{sec:results_no_drift}}

This section examines the effect of coagulation and fragmentation between dust grains and non-drifting (vertically settled) pebbles on the density structure (model~C+D+Co; row three in Table~\ref{tab:grid}). While the previous models already included pebbles, they did not participate in any evolution. In model~C+D+Co, we start with start with ice on pebbles as well instead of the dry pebbles initialised earlier. Ices are distributed iteratively by (1) calculating the vapour to ice on dust grain equilibrium (Eq.~\ref{eq:ice_vapor}), (2) redistributing the ices between pebbles and dust grains to enforce the same surface density ratio as seen between dust grains and pebbles, and (3) repeating this process to an equilibrium state.

The C+D+Co model is then evolved using chemistry, dynamics (advection and diffusion), coagulation and fragmentation, but crucially without radial movement of pebbles (no pebble drift and 1D radial diffusion). 

\begin{figure*}
    \centering
    \includegraphics[width=1.0\linewidth]{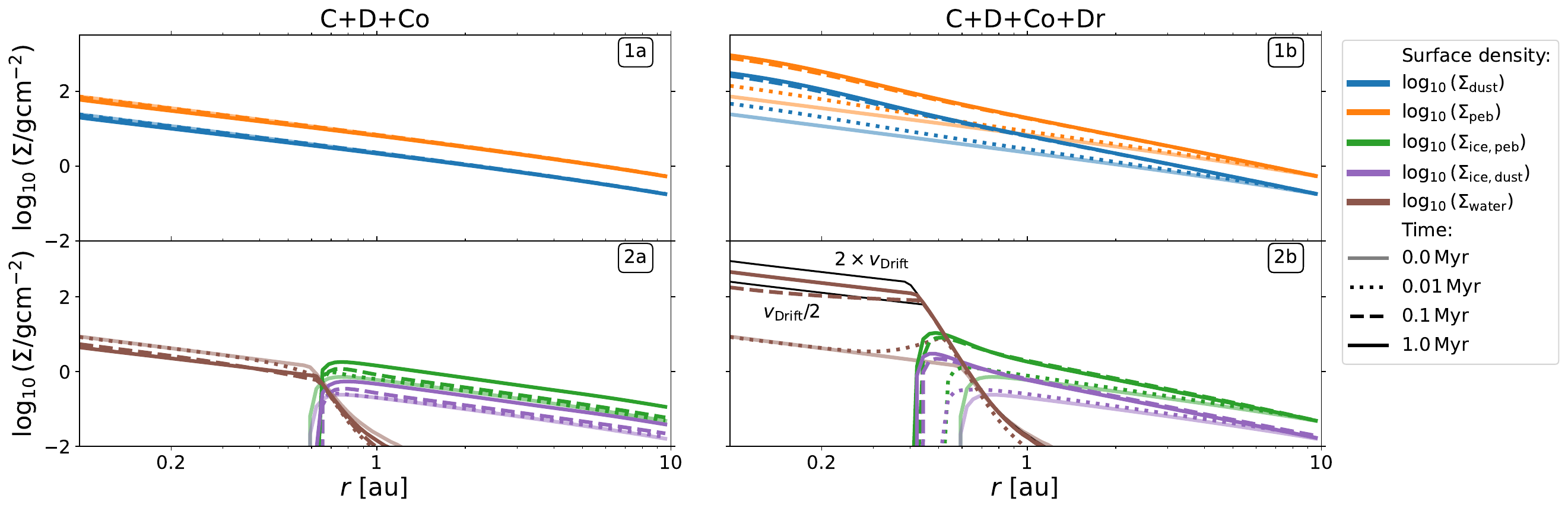}
    \caption{Evolution of surface densities for model~C+D+Co (left panels) and model~C+D+Co+Dr (right panels). The different colours correspond to different surface densities with the dust, pebble, ice on pebbles, ice on dust grains, and water vapour surface density depicted in blue, orange, green, purple, and brown, respectively. The upper panels show all solids while the lower panels depict all water surface densities. The initial conditions (faint lines) are the same for all models. The surface densities at $0.01\,\rm Myr$ (dotted line), $0.1\,\rm Myr$ (dashed line) and $1\,\rm Myr$ (solid line) are displayed as well. The black lines in panel~2b depict the end state water surface density for the supporting models with half and double the fiducial drift velocity (see Table~\ref{tab:grid}).}
    \label{fig:surface_dens}
\end{figure*}

The evolution of the surface densities is shown in panels~a of Fig.~\ref{fig:surface_dens}. The surface density of pebbles and dust grains does only change marginally over time owing to the fact that the simulation is initialised with an equilibrium distribution (panel~1a). As explained in Sect.~\ref{sec:pebbles}, we assume a dust to pebble surface density ratio of $1/3$ (Eq.~\ref{eq:fcol}). The only marginal temporal changes are due to the 2D advection velocity field that affects dust grains leading to a minimal decrease of both dust and pebble surface density at small radii (as tentatively seen in panel~1a of Fig.~\ref{fig:surface_dens}). 

\begin{figure*}
    \centering
    \includegraphics[width=1.0\linewidth]{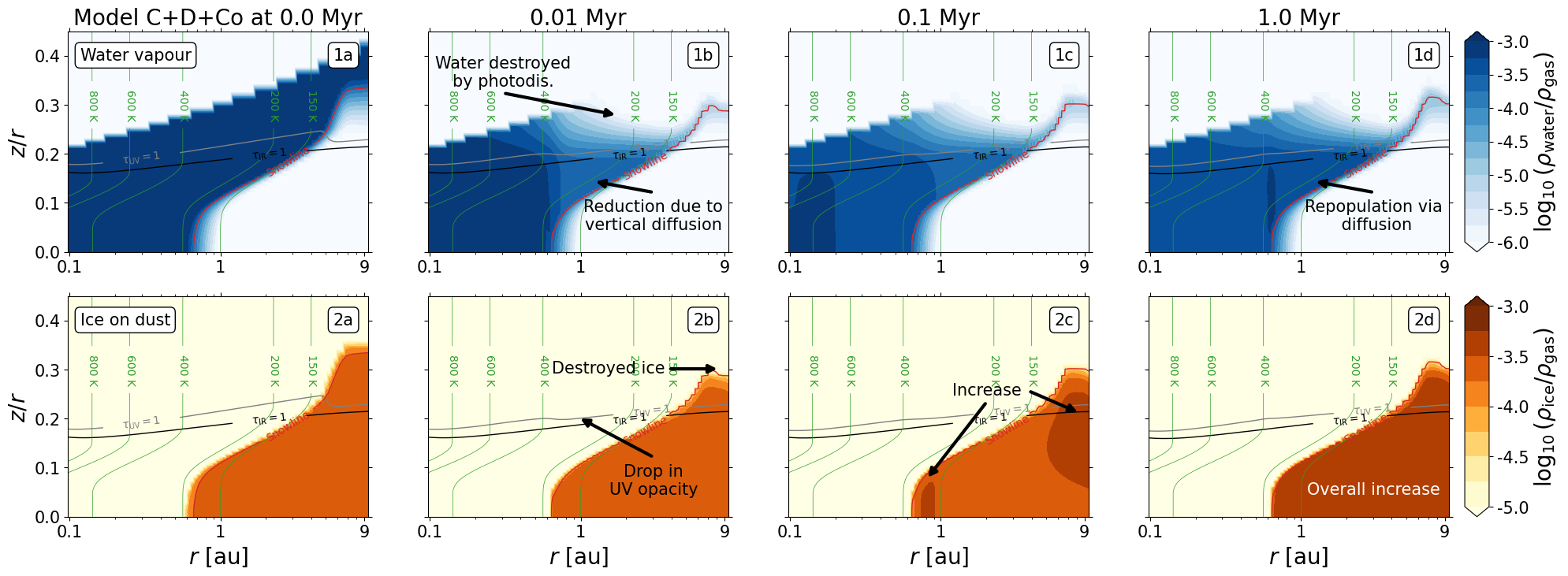}
    \caption{The dynamical evolution of water in the model~C+D+Co (row three of Table~\ref{tab:grid}). Every column depicts a different time during the evolution (initial conditions, $0.01\,\rm Myr$, $0.1\,\rm Myr$, and $1\,\rm Myr$) of the water vapour concentration (top row),  water ice concentration (last row). The snowline, temperature contours and opacity lines are shown for reference.}
    \label{fig:pebble_nodrift_2d}
\end{figure*}

Both ice on pebbles and ice on dust grains show an increase in surface density outside of $\sim0.6\,\rm au$ (roughly the midplane snowline) with a strong depletion just inside of that. The corresponding 2D ice on dust grain concentration can be seen in panel~2a-2d of Fig.~\ref{fig:pebble_nodrift_2d}. Ices in the surface are quickly removed due to photodissociation of water vapour (as already seen in Fig.~\ref{fig:chem_and_dyna_2d}). The increase of ice on dust grain concentration (e.g. panel~2c and 2d of Fig.~\ref{fig:pebble_nodrift_2d}) is due to cold water formation and subsequent condensation (as discussed in Sect.~\ref{sec:results_chem}). 
Additionally, the lower concentration of ice on dust grains compared to water vapour inside the snowline\footnote{This is a consequence of the initialisation with part of the ice being captured on pebbles.} leads to a radial outwards (cold finger effect) and vertical downwards diffusion of water vapour which increases the ice on dust grain concentration over timescales of $\sim1\,\rm Myr$ (panel~2d).

The water vapour concentration in the cold disc surface decreases quickly due to photodissociation (e.g. panel~1b of Fig.~\ref{fig:pebble_nodrift_2d}) as already seen in the previous models, which leads to a drop in UV opacity (panel~2b). Additionally, vapour above the ice region but below the $\tau_{\rm UV}=1$ line is depleted. This is due to the vertical cold finger effect, with ice being captured by pebbles in the midplane leading to a vertical downwards diffusion of vapour in the surface \citep[e.g.][]{Meijerin2009,Krijt2016,Du2017}. After $1\,\rm Myr$ of evolution, radial diffusion minimised the radial concentration gradient and repopulates this depleted region (panel~1d of Fig.~\ref{fig:pebble_nodrift_2d}).
This can be seen as well in panel~2a of Fig.~\ref{fig:surface_dens}, where the water vapour surface density is initially decreasing outside the snowline but increases slowly between $0.01-1\,\rm Myr$. This comes with the consequent decrease in vapour surface density inside the snowline.

To summarise, adding coagulation and fragmentation of dust grains and non-drifting pebbles to the simulation (1) leads to the a reduction of water vapour above the ice reservoir due to the vertical cold finger effect, which (2) is replenished by radial diffusion on longer ($\sim1\,\rm Myr$) timescales, leading to (3) a decrease of water surface density inside the midplane snowline.

\subsection{Effect of pebble drift on water distributions (model~C+D+Co+Dr)\label{sec:results_drift}}

\begin{figure*}
    \centering
    \includegraphics[width=1.0\linewidth]{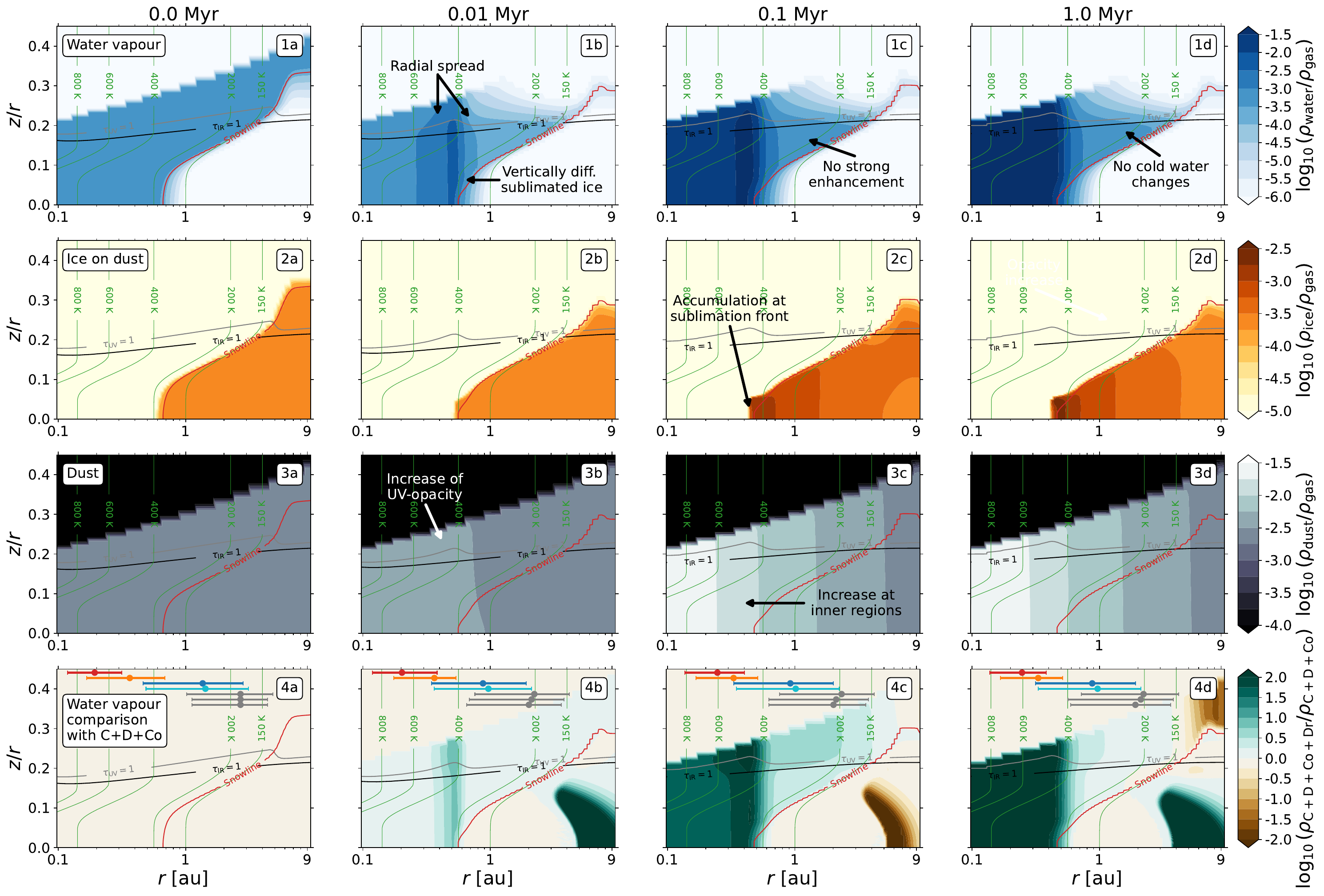}
    \caption{The dynamical evolution of water in the full model~C+D+Co+Dr (fourth row of Table~\ref{tab:grid}). Every column depicts a different time during the evolution (initial conditions,  $0.01\,\rm Myr$, $0.1\,\rm Myr$, and $1\,\rm Myr$) of the water vapour concentration (top row), water ice concentration (second row), dust concentration (third row). The last row shows the ratio in water vapour concentration between model~C+D+Co+Dr and model~C+D+Co. The horizontal lines in panels~4 show (top to bottom) the emitting regions in model~C+D+Co+Dr of the diagnostic water lines (more details in Sect.~\ref{sec:results_spec}) and far-IR lines at $40.69\,\rm \mu m$ ($E_{\rm U}=550\,\rm K$), at $46.48\,\rm \mu m$ ($E_{\rm U}=410\,\rm K$), and at $61.81\,\rm \mu m$ ($E_{\rm U}=552\,\rm K$). The snowline, temperature contours and opacity lines are shown for reference.}
    \label{fig:pebble_full_2d}
\end{figure*}

The full model~C+D+Co+Dr (row four of Table~\ref{tab:grid}) adds pebble drift and pebble diffusion to the previously discussed chemical, diffusion/advection, and coagulation/fragmentation processes.

The influx of pebbles at the outer edge of the simulations matches the drift flux from the outermost to the adjacent cell at the beginning of the simulation and is kept constant over time (details in Sect.~\ref{sec:pebbles}). The assumptions lead to a pebble flux of $\sim 2.4\times 10^{-4}\,\rm M_{\oplus}/yr$, which is similar to values derived from cold water lines in JWST/MIRI spectra \citep[e.g.][]{Romero-Mirza2024,Krijt2025}, 1D dust evolution models \citep[e.g.][]{Birnstiel2012,Drazkoska2021}, and the pebble flux needed to explain dust and CO observations of IM\,Lup \citep[][]{Bosman2023}. We choose to keep this influx from the outer disc constant throughout the simulation for simplicity\footnote{A constant pebble flux based on the initial drift velocity in the outer most cell results after $1\,\rm Myr$ in a total accreted pebble mass ($\sim 6.5\times 10^{-4}\,\rm M_\odot$) that exceeds the available solid mass of the disc ($2\times 10^{-4}\,\rm M_\odot$). However, we see that the simulation establishes a quasi-equilibrium already at about $0.2\,\rm Myr$ at which point the accreted mass has not yet exceeded the total available reservoir. Therefore, the $1\,\rm Myr$ snapshots can be seen as the drift equilibrium case where only chemical water formation alters the water density structure.}, and leave the implementation of more complex time dependent pebble flux profiles to future studies.

The surface density evolution of model~C+D+Co+Dr can be seen in the panel~1b and 2b of Fig.~\ref{fig:surface_dens}. First of all, we note that the inward drift of pebbles increases the surface density of pebbles at small radii. This leads to a steeper slope of the pebble surface density matching (within $0.5\,\%$) the analytic prediction of $\Sigma_{\mathrm{P}}\propto r^{-0.75}$ in the fragmentation limited case \citep{Birnstiel2012}. While the increase in pebble surface density is a direct consequence of drift, the dust grain surface density increases as well due to the constant fragmentation of pebbles. Fig.~\ref{fig:pebble_full_2d} shows the 2D water vapour concentration, ice on dust grains concentration, and dust concentration. Panels~3a-3d show how the dust concentration increases at small radii (e.g. panel~3c). The increase in dust concentration leads to a small increase in dust opacity in the inner disc. Generally, no strong vertical gradients exists in the dust concentration. Only at $0.01\,\rm Myr$ (panel~3b) a vertical gradient is visible around $1\,\rm au$. This confirms that vertical diffusion acts very quickly in minimising concentration gradients in the inner disc \citep[see e.g. Fig~2 in][]{Semenov2011}.

The water vapour concentration (panels~1a-1d of Fig.~\ref{fig:pebble_full_2d}) shows strong changes compared to the already discussed chemical and dynamical effects. In-drifting pebbles release their water vapour inside the snowline (panel 1b), which quickly diffuses vertically upwards in the observable surface layer. In the regions with most water enhancement, the UV opacity increases significantly (e.g. panel~3b) due to water UV shielding \citep[e.g.][]{Bosman2022}. Through radial diffusion this concentration peak is spreading radially as well (e.g. panel~1b), with the disc region inside the water snowline reaching a nearly constant concentration after $1\,\rm Myr$ (panel~1c). This new concentration is about $50-60$ its initial value (see panel~2b of Fig.~\ref{fig:surface_dens}), consistent with enhancements found using 1D models \citep[see e.g. Fig. 6 in ][]{Schneider2021}. \cite{Cuzzi2004} show using simplified 1D model that for constant pebble flux, the steady state enhancement factor inside the snowline can be approximates as $E_{\rm increase}\approx2f_{\mathrm{L}}/(3\alpha)$, with $f_{\mathrm{L}}$ denoting the mass ratio between all large and small solids ($1/3$ in our models) and $\alpha$ denoting the $\alpha$ parameter ($10^{-3}$ in model~C+D+Co+Dr). This equation predicts an enhancement factor of $\sim200$, which is only a factor of $\sim 4$ larger than the actual enhancement of $50-60$. The difference might originate from several model differences, including the different between 1D and 2D models. Additionally, \cite{Cuzzi2004} approximate the timescale on which constant enhancement is reached as $(40/\alpha)$ orbital periods. Using the orbital period at the midplane snowline this estimate of $\sim0.05\,\rm Myr$ matches the actual timescale which is between $0.01-0.1\,\rm Myr$ (see panel~2b of Fig.~\ref{fig:surface_dens}). While the region inside the water snowline shows a constant enhancement, the outward movement of vapour is much slower. After $0.1\,\rm Myr$, the concentration above the ice region below the photodissociation region has a water concentration comparable to the initial value (compare panel~1a and 1c of Fig.~\ref{fig:pebble_full_2d}), recovering from the slight drop due to the vertical cold-finger effect (panel~1b). Even at $1\,\rm Myr$ (panel~1d) the water concentration does not increase significantly. This means that there is no strong enhancement in water concentration even though the water surface density inside the snowline has increased by a factor of $50-60$.  

The influx of icy pebbles leads to a buildup of ice on dust grains as well. Ice on dust accumulates at the sublimation front (panel~2c of Fig.~\ref{fig:pebble_full_2d} and panel~2b of Fig.~\ref{fig:surface_dens}) and shows the same radial slope as the dust and pebble surface density. This means that more and more ice on dust grains exists inside the initial snowline (defined as equal vapour and ice concentration) due to the fact that the saturation concentration is reached and extra water remains as ices.
The ice on pebble surface density (panel~2b of Fig.~\ref{fig:surface_dens}) shows the same behaviour due to the constant exchange between dust and pebbles via coagulation and fragmentation.

The influence of pebble drift on water vapour can be seen in panels~4 of Fig.~\ref{fig:pebble_full_2d}, which compares the 2D water concentration of two identical models that only differ in their inclusion (model~C+D+Co+Dr) and exclusion (model~C+D+Co) of pebble drift. After $0.01\,\rm Myr$ the vertically well mixed plume of water at the water snowline enhances the water concentration in the drift model compared to model~C+D+Co by factors of $~10$ (panel~4b). This enhancement spreads inwards leading to a water concentration inside the water snowline about a factor of $60$ higher in the drift case compared to the no drift case at $1\,\rm Myr$ (panel~4d). However, above the snow surface the concentrations look very similar with changes of less than $0.25\,\rm dex$ (panel~4d). Only the shift of the snowline at the outermost region leads to a $\sim 1\,\rm dex$ change in water concentration. We discuss the observational signatures of this behaviour in Sect.~\ref{sec:results_spec}.

\begin{figure*}
    \centering
    \includegraphics[width=1.0\linewidth]{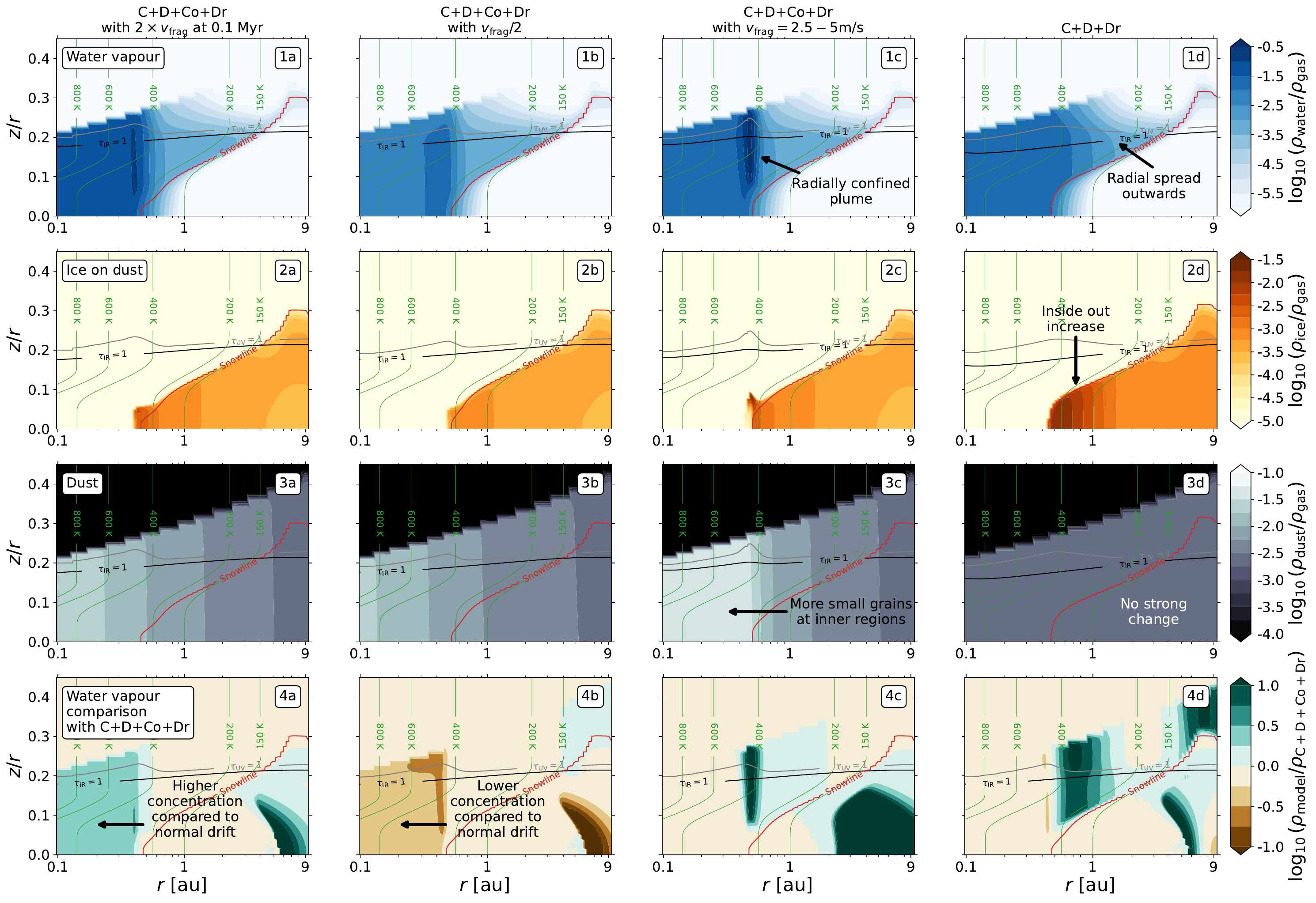}
    \caption{The dynamical evolution in all support models including diverse pebble drift scenarios (see Table~\ref{tab:grid}). Every column depicts a different different model at $0.1\,\rm Myr$ with panels~1 showing model~C+D+Co+Dr with $2 \times v_{\mathrm{frag}}$, panels~2 showing model~C+D+Co+Dr with $v_{\mathrm{frag}}/2$, panels~3 showing model~C+D+Co+Dr with $v_{\mathrm{frag}}=2.5-5\,\rm m/s$, and panels~4 showing model~C+D+Dr. The water vapour concentration (top row), water ice concentration (second row), dust concentration (third row) are displayed. The last row shows the ratio in water vapour concentration between the column's model and model~C+D+Co+Dr. The snowline, temperature contours and opacity lines are shown for reference, with all contours in the row~d taken from model~C+D+Co+Dr.}
    \label{fig:pebble_support_2d}
\end{figure*}

\subsubsection{The effect of different pebble fluxes\label{sec:results_vfrag}}

To test the effect of different pebble fluxes, we set up two supporting models with half and double the drift velocity (Table~\ref{tab:grid}). Since the pebble influx is linked to the drift velocity, this results in pebble fluxes of $\sim 1.2\times 10^{-4}\,\rm M_{\oplus}/yr$ and $\sim 4.8\times 10^{-4}\,\rm M_{\oplus}/yr$, respectively. The 2D water concentration, ice on dust grain concentration, and small grain concentration at $0.1\,\rm Myr$ can be seen in panels~a and b of Fig.~\ref{fig:pebble_support_2d}. For double and half the fragmentation velocity compared to model~C+D+Co+Dr the water concentration inside the midplane snowline is enhanced and reduced, respectively (panels~4). This is also shown by the quasi-steady water vapour surface densities (at $1\,\rm Myr$) for these two models in panel~2b of Fig.~\ref{fig:surface_dens}. In the inner disc region ($r\lesssim0.5\,\rm au$)the water surface densities decreases (for $v_{\mathrm{Drift}}/2$) and increases (for $2\times v_{\mathrm{Drift}}$) by a factor of $2$, confirming the suggested linear relation between enhancement inside the midplane snowline and pebble flux \citep{Cuzzi2004}. However, the surface density profiles outside of $\sim0.5\,\rm au$ are nearly identical (as also shown by panels~4a and 4b of Fig.~\ref{fig:pebble_support_2d}), showing that the lack of water vapour increase outside the midplane snowline is independent in our setup of the assumed drift velocity and pebble flux. This is a somewhat surprising result and we briefly explore why this behaviour is seen.

\subsubsection{The pebble conveyor belt effect\label{sec:results_noexchange}}

Assuming that the vertical diffusion is faster than the radial gas transport, sublimated water will be well mixed vertically on the radial diffusion timescale. This means that water vapour that moves outside the midplane snowline will (at least partly) freeze out as ice on dust grains. These grains can coagulate in the midplane transferring the ice onto pebbles. Additionally, vapour can also directly freeze out on pebbles in the midplane if it reaches that region. Due to the constant and fast inwards drift of pebbles (compared to radial outwards diffusion), water is effectively moved back into warmer regions where it will sublimate again. Therefore, outwards diffusing vapour in the disc surface is trapped in a cycle that reduces its efficiency. This effect is reminiscent of small grains becoming trapped in the midplane or pressure bumps if the collision timescale is shorter than the diffusion timescale \citep[][]{Krijt2016a,Yang2025}.

To test this scenario, we evolve model~C+D+Dr (see Table~\ref{tab:grid}), which mimics model~C+D+Co+Dr in every aspect but does not allow for coagulation and fragmentation, effectively limiting the ability of dust and pebbles to exchange ice reservoirs. Physically, this scenario could correspond to a picture where dust coagulation is halted by bouncing rather than fragmentation \citep[e.g.][]{Zsom2010,Dominik2024}.
The 2D concentration structure of water vapour, ice on dust grains, and dust concentration at $0.1\,\rm Myr$ is shown in panels~d of Fig.~\ref{fig:pebble_support_2d}. Similarly to model~C+D+Co+Dr, pebbles sublimate inside the snowline leading to an enhancement of water vapour. However, not allowing for coagulation breaks the aforementioned cycle, which has been seen in model~C+D+Co+Dr. Therefore, efficient outwards diffusion can be seen for vapour (panel~1d) and ice on dust grains (panel~2d). This leads to efficient concentration enhancement of cold water above the ice reservoir and a corresponding strong increase in ice on dust grain concentration. 
This shows that the cycle including coagulation and pebble drift is responsible for the lack of outwards diffusion in model~C+D+Co+Dr. Therefore, it likely matters if dust is models with discrete size bins or as full size distributions.

\subsubsection{The effect of changing fragmentation velocity}
\label{sec:results_changevfrag}

Lastly, we explore the effect of changing dust properties for dry and wet pebbles. Several studies highlight the change in fragmentation velocity for pebbles if they are dry or covered by an icy mantel \citep[e.g.][]{Dominik1997,Gundlach2015}. As shown by \cite{houge_smuggling_2025} a decrease in fragmentation velocity inside the water snowline leads to an increase in the concentration of small grains which in turn increases the dust opacity in this region. Therefore, \cite{houge_smuggling_2025} stress that this effect leads to a constant observable water column density despite ongoing pebble drift.

We explore this scenario by halving the fragmentation velocity (from $5\,\rm m/s$ to $2.5\,\rm m/s$) at radii where ratio between ice on pebble and pebble surface density exceeds $0.01$. We note that this change by a factor of $2$ is quite conservative with \cite{houge_smuggling_2025} exploring changes by a factor of $10$. However, we leave a full exploration on different factors to future studies and focus on the qualitative results.

The model concentrations at $0.1\,\rm Myr$ are shown in panels~c of Fig.~\ref{fig:pebble_support_2d}. The plume of water is radially more confined compared to model~C+D+Co+Dr (panel~1c of Fig.~\ref{fig:pebble_full_2d}). This is due to the change in drift velocity which is connected to the fragmentation velocity via Eq.~\ref{eq:drift} and Eq.~\ref{eq:stokes_peb}. According to these equations dry pebbles drift slower than wet pebbles leading to a traffic jam effect at the water snowline. Therefore, the radial region in which ice sublimates gets more confined. As a consequence, the water concentration above this region becomes enhances (panel~4c of Fig.~\ref{fig:pebble_support_2d}).

Another consequence of the changing fragmentation velocity is an increase in the dust/pebble ratio. The surface density ratio changes proportional to the inverse square root of the fragmentation velocity ratio according to Eq.~\ref{eq:fcol}. Additionally, the aforementioned traffic jam leads to a sharp change in the surface density profiles in both dust and pebbles (Fig.~\ref{fig:support_surface_dens}).
We explore the effect on the observable water reservoir in Sect.~\ref{sec:results_spec}.

We conclude that drift of icy pebbles in combination with chemistry, diffusion and advection, and coagulation and fragmentation results (for our choice of $\alpha$ and pebble properties) in (1) an enhancement of water vapour inside the water snowline via sublimating pebbles, which is vertically distributed on short timescales, and (2) radial transport processes spreading the water plume mostly inwards, which therefore results in little enhancement of the observable cold water region compared to an identical model without drift. This is surprising given the proposed observational relation between the strength of cold water excess and pebble drift. We also note that (3) the efficiency of radial outwards transport of water is hampered by a cycle of radial and vertical diffusion, freeze out, coagulation, and inwards drift, (4) the UV opacity increasing due to the increase in water vapour with an small increase of the IR opacity in the inner regions due to the raise of dust grains concentration, and (5) that changes in the fragmentation velocity for dry and wet pebbles confine the water radially and enhance the dust concentration inside the midplane snowline.

\section{Simulated mid-IR spectra}
\label{sec:results_spec}

\begin{figure*}
    \centering
    \includegraphics[width=0.99\linewidth]{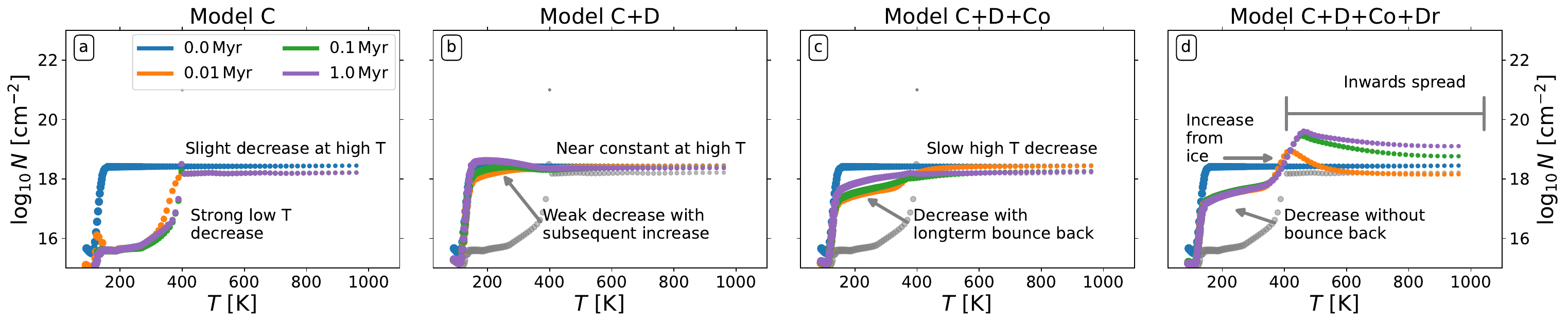}
    \caption{Slab conditions extracted from the chemistry only model (left panels) and model including chemistry and diffusion (second panels), the model without drift (third panels), and the full model (right panels) at $0.0\,\rm Myr$ (blue), $0.01\,\rm Myr$ (orange), $0.1\,\rm Myr$ (green), and $1.0\,\rm Myr$ (purple). Every marker represents a single column of the 2D density structure turned into input for a slab model (column density above $\tau_{\mathrm{IR}}=1$ and mass average temperature in that visible column). The marker size is proportional to the logarithm of the slab's emitting area. The grey dots in panel~b-d are model~C at $1\,\rm Myr$.}
    \label{fig:conditions}
\end{figure*}

\begin{figure*}
    \centering
    \includegraphics[width=1.0\linewidth]{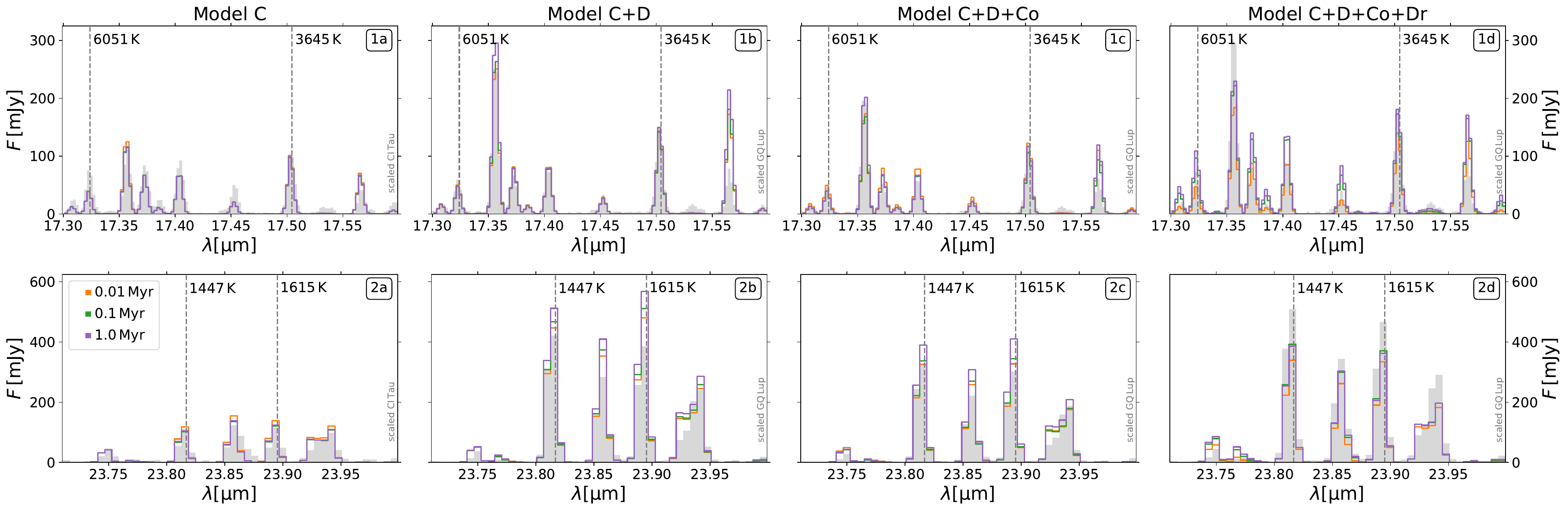}
    \caption{Spectral zoom-ins for the chemistry only model (left panels) and model including chemistry and diffusion (second panels), the model without drift (third panels), and the full model (right panels) at $0.01\,\rm Myr$ (orange), $0.1\,\rm Myr$ (green), and $1.0\,\rm Myr$ (purple). Overplotted are the JWST/MIRI spectra of CI\,Tau (panel~1a and 2a) and GQ\,Lup (all other panels) scaled to the peak flux of the $E_{\mathrm{U}}=3645\,\rm K$ water line of the $1\,\rm Myr$ model for comparison (details in Sect.~\ref{sec:discus_observ}). The zoom-in from $17.29\,\rm \mu m$ to $17.6\,\rm \mu m$ (top row) contains lines that are used to analyse hot water ($\sim 17.32\,\rm \mu m$) and warm water ($\sim 17.5\,\rm \mu m$) marked with their upper level energy. The zoom-in from $23.71\,\rm \mu m$ to $24\,\rm \mu m$ (bottom row) contains two marked cold water lines.}
    \label{fig:specs}
\end{figure*}

\begin{figure}
    \centering
    \includegraphics[width=1.0\linewidth]{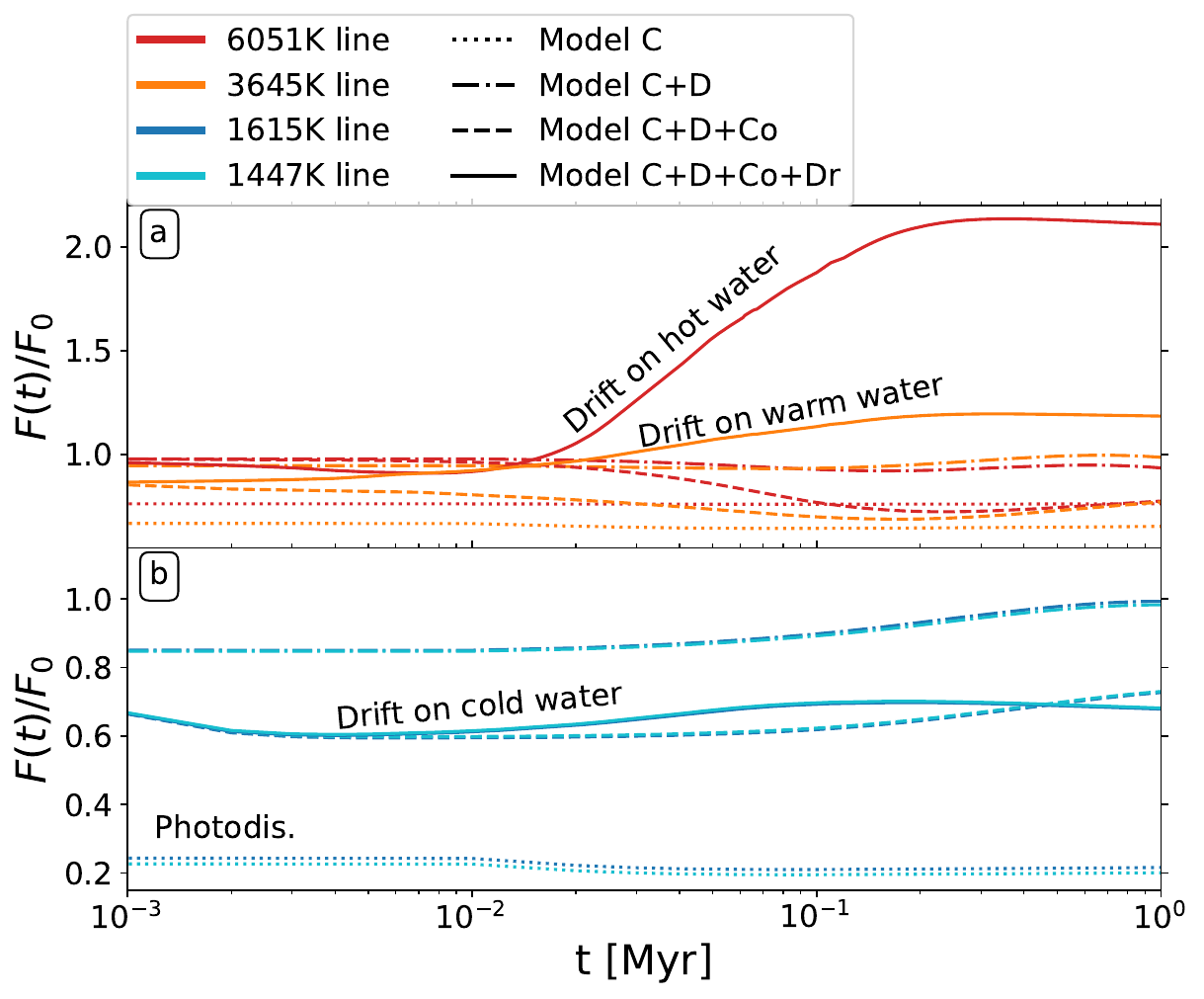}
    \caption{Evolution of the water emission line strength (compared to the initial value) of the diagnostic hot, warm (both in panel~a), and cold water lines (panel~b) for model~C (dotted; showing clear signs of photodissociation in the cold water line strengths of $40\,\%$ of their initial value), model~C+D (dashed dotted), model~C+D+Co (dashed), and full model~C+D+Co+Dr (solid lines; indicating the effects of drift on all water line strengths).}
    \label{fig:line_strength}
\end{figure}

The 2D density structures of all models are used to simulate mid-IR spectra. This means that the changes in water density and dust density can be directly linked to observable differences in JWST/MIRI spectra. For doing so, every modelled column is approximated by a slab model that accounts for the columns emitting area, observable column density in the IR, and the mass average temperature (see Fig.~\ref{fig:slab_extraction} and Sect.~\ref{sec:sim_obs} for details). The extracted emitting conditions at the initial conditions, $0.01\,\rm Myr$, $0.1\,\rm Myr$, and $1\,\rm Myr$ for all four main models are displayed in Fig.~\ref{fig:conditions}.

At the initial conditions (blue markers in all panels), the emitting column density above the dust optical surface is $\sim10^{18.4}\,\rm cm^{-2}$ for all $T\gtrsim150\,\rm K$ with a steep drop to column densities as low as $10^{15.5}\,\rm cm^{-2}$ at lower temperatures. The transitioning temperature of $\sim 150\,\rm K$ corresponds to the temperature where the snowline crosses the $\tau_{\rm IR}=1$ line. Due to the constant initial water and dust concentration both exhibit the same vertical distribution which leads to a radially constant column density if the snowline is below the $\tau_{\rm IR}=1$ line. At lower temperatures, large fractions of water in the observable region are frozen, reducing the vapour column density.

\subsection{Effect of chemical processing on mid-IR spectra (model~C)}

Model~C (panel~a of Fig.~\ref{fig:conditions}) shows how chemistry leads to a strong column density decrease at low temperatures ($T<400\,\rm K$), with only a small decrease of $<0.3\,\rm dex$ at high temperatures ($T>400\,\rm K$). This decrease is due to photodissociation destroying water in the disc surface (see Sect.~\ref{sec:results_chem}) and therefore dependent on the assumed photodissociation timescale and UV opacity. Water self-shielding ensures the survival of a layer of water between the $\tau_{\rm UV}=1$ and $\tau_{\rm IR}=1$ line at $T>400\,\rm K$. Since the observable column density is dominated by the regions just above the $\tau_{\rm IR}=1$ line, photodissociation only leads to a small drop in warm and hot water column density. On the other hand, photodissociation at $T<400\,\rm K$ transforms water into oxygen which does not shield water from UV radiation, lowering the $\tau_{\rm UV}=1$ line. Therefore, the column density of water at these temperatures is set by the thin layer of water between the dust $\tau_{\rm UV}=1$ and $\tau_{\rm IR}=1$ line and the low abundance water region in the surface (see e.g. panel~1c of Fig.~\ref{fig:chem_2d}) which explains the decrease in column density of about $3\,\rm dex$. 

Using these conditions, JWST/MIRI water spetra are simulated. Zoom-ins to the four water lines that are commonly used to characterise the water temperature distribution \citep[introduced by][]{Banzatti2025} are shown in panel~1a and 2a of Fig.~\ref{fig:specs}. The spectra after $0.01\,\rm Myr$ (orange), $0.1\,\rm Myr$ (green), and $1\,\rm Myr$ (purple) are displayed, with the hot (marked by its upper level energy $E_\mathrm{u}$ of $6051\,\rm K$), warm ($E_\mathrm{u}=3645\,\rm K$), and two cold ($E_\mathrm{u}=1447\,\rm K$ and $E_\mathrm{u}=1615\,\rm K$) being labelled. The two cold water lines (panel~2a) have a similar strength to the line between them (at $23.85\,\rm \mu m$, with $E_{\mathrm{U}}=2892\,\rm K$). This pattern has already been seen by \cite{Vlasblom2025} using full thermochemical models and identified as no strong cold water enhancement. 

The line strength evolution of the diagnostics water lines (indicated by their colour) normalised to their initial fluxes is shown in Fig.~\ref{fig:line_strength} for all models (indicated by the line style). The hot water line stays nearly constant over time with about $75\,\%$ of its initial flux (panel~1a in Fig.~\ref{fig:specs} and dotted red line in panel~a of Fig.~\ref{fig:line_strength}). The warm water line retrains about $65\,\%$ of its initial strength (seen by the nearly horizontal dotted orange line in panel~a of Fig.~\ref{fig:line_strength} at $0.65$). The two cold water lines loose $\sim75\,\%$ of their flux (panel~b of Fig.~\ref{fig:line_strength}) in line with the steep drop in column density at $T<400\,\rm K$. This (initial condition depended) drop happens already before $0.001\,\rm Myr$ and is not followed by any other later major flux changes.

\subsection{Effect of diffusion and advection on mid-IR spectra (model~C+D)}
\label{sec:effect_diff}

Accounting for diffusion and advection together with the chemical setup (model~C+D) leads to very different emission conditions (panel~b of Fig.~\ref{fig:conditions}). The column density decrease at $0.01\,\rm Myr$ at $T<400\,\rm K$ due to photodissociation is much weaker (less than $0.4\,\rm dex$) compared to model~C with a subsequent increase of column density to $\sim10^{18.7}\,\rm cm^{-2}$ beyond its initial value at $0.1,\rm Myr$ and $1\,\rm Myr$. This means that the cold water lines (panel~2b of Fig.~\ref{fig:specs}) increase over time.  This is underpinned by Fig.~\ref{fig:line_strength} showing the cold water line flux dropping to $85\,\%$ of its initial value before increasing back to it at $1\,\rm Myr$ (panel~b). The initial conditions with constant water abundances already show spectral characteristics of a strong cold water excess (details in Sect.~\ref{sec:discus_observ}). This should be kept in mind, when interpreting the increase in model~C+D past these conditions. As discussed in Sect.~\ref{sec:results_alpha}, stronger UV fields would decrease the water concentration in the surface. Therefore, adding diffusion to a static chemical model does not necessarily lead to an increase beyond the initial column densities. However, even in a strong photodissociation case model~C+D would show higher observable cold water column densities than an equivalent model~C if the Damköhler number at $\tau_{\rm IR}$ is not significantly larger than $1$. A similar behaviour can be seen for the warm water line, which increases stays close to its initial value (panel~a). At high temperatures the column densities stay nearly constant, with a slight decrease at $1\,\rm Myr$ (panel~b of Fig.~\ref{fig:conditions}) due to radial diffusion of vapour into colder regions. This is reflected in the hot water line (panel~1b of Fig.~\ref{fig:specs} and panel~a of Fig~\ref{fig:line_strength}) with the line dropping in flux by only about $5,\%$ and staying nearly constant between $0.001-1\,\rm Myr$.

\subsection{Effect of non-drifting pebbles on mid-IR spectra (model~C+D+Co)}

For model~C+D+Co, photodissociation and the vertical cold finger effect lead to a depletion (of about $1\,\rm dex$) in observable column density below $400\,\rm K$ at $0.01\,\rm Myr$ (panel~c of Fig.~\ref{fig:conditions}). Over time radial diffusion replenishes this reservoir resulting in column densities of about $10^{18.0}\,\rm cm^{-2}$ at $T<400\,\rm K$ and subsequent decrease of $0.4\,\rm dex$ at $T>400\,\rm K$. This is mirrored in the spectrum which shows a slow decrease in hot water line flux (panel~1c of Fig.~\ref{fig:specs}) with the initial flux being retrained till about $0.01\,\rm Myr$, but dropping afterwards to about $75\,\%$ of that value. The warm water line which is more sensitive to the cold reservoir compared to the hot water lines drops by about $30\,\%$ of flux at $0.1\,\rm Myr$, with a subsequent increase to $75\,\%$. The cold water lines (panel~2c of Fig.~\ref{fig:specs}) decrease to about $60\,\%$ of their initial flux at $0.01\,\rm Myr$, but rise again to more than $70\,\%$ at $1\,\rm Myr$ (panel~b of Fig.~\ref{fig:line_strength}). 

The impact of adding pebbles can be seen when comparing model~C+D+Co with model~C+D in Fig.~\ref{fig:line_strength}. The warm (panel~a) and cold water lines (panel~b) evolve similarly with the lines of model~C+D+Co $\sim15\,\%$ and $\sim25\,\%$ weaker, respectively. This indicates how part of the water reservoir above snow region is captured due to the vertical cold finger effect. The hot water line (panel~a) shows a similar strength for model~C+D+Co and model~C+D up to $0.01\,\rm Myr$, with the hot water line of model~C+D+Co decreasing afterwards due to stronger radial diffusion replenishing the depleted outer reservoir. Even though, the cold water lines in model~C+D+Co are weaker than in model~C+D, they still account for a significant increase compared to model~C.

\subsection{Effect of pebble drift on mid-IR spectra (model~C+D+Co+Dr)}

Lastly, we examine the effect of pebble drift on observables (model~C+D+Co+Dr). As seen in panel~d of Fig.~\ref{fig:conditions}, the sublimation of icy pebbles inside the water snowline leads to an increase in observable column density to up to $10^{19.0}\,\rm cm^{-2}$ around $T=350-500\,\rm K$ at $0.01\,\rm Myr$. This temperature range corresponds to temperature above the $\tau_{\rm IR}=1$ line vertically above the sublimation region of pebbles inside the midplane. This increase in column density coincides with the already discussed decrease at lower temperature ($T<400\,\rm K$). The observable column density shifts inwards to higher temperatures over time with the peak of $10^{19.5}\,\rm cm^{-2}$ at a temperature of $460\,\rm K$ after $1\,\rm Myr$. At $0.01\,\rm Myr$ the column density at temperature higher than $530\,\rm K$ drop below their initial value due to an increase in IR opacity resulting from the increase in dust concentration (e.g. panel~3b of Fig.~\ref{fig:pebble_full_2d}). However, at $0.1\,\rm Myr$ the column densities of all grid points with temperatures above $400\,\rm K$ increases past their initial value (panel~d of Fig.~\ref{fig:conditions}). We note that the column densities in model~C+D+Co+Dr are slightly higher than typically retrieved values from JWST/MIRI observations of $\sim10^{18}-10^{19}\,\rm cm^{-2}$ \citep[e.g.][]{Romero-Mirza2024,Temmink2024} even though some extreme objects show higher column densities \citep[e.g. HD\,35929;][]{Kaeufer2026}. While at these high column densities opacity effects become important, we show that the general conclusions are consistent with higher and lower pebble fluxes with the latter resulting in column densities closer to observed values (Fig.~\ref{fig:support_conditions}). Alternatively, changes in the photodissociation rate or UV and IR opacity of the dust could explain mismatches between our models and observations. To fully explore their impact, evolving temperature structures and radiative transfer simulations of the interstellar and stellar UV are needed which is beyond the scope of this work.

Interestingly, the column density at $T<350\,\rm K$ decreases compared to their initial values as already discussed in the context of model~C+D+Co but in contrast to this model does not bounce back to higher values. This means that the column densities at these temperatures are below values of model~C+D but still significantly higher than those of model~C. The evolution of the emitting conditions results in changes of the observed spectrum (panel~1d and 2d of Fig.~\ref{fig:specs}). Panel~a of Fig.~\ref{fig:line_strength} shows that the hot and warm water line increase by more than a factor of $2$ and $20\,\rm \%$ in strength, respectively. The cold water lines (panel~b of Fig.~\ref{fig:line_strength} show the same initial drop in flux as model~C+D+Co, but recover to about $70\,\%$ of their initial flux at $0.1\,\rm Myr$ with drop by a few percent later on. 

To summarise, (1) chemical processing through photodissociation weakens the cold water lines most significantly, (2) adding diffusion minimises this effect, with the cold water line fluxes increasing on a $\sim 1\,\rm Myr$ timescale via radial diffusion for the assumed photodissociation rate, (3) non-drifting pebbles capture part of the observable cold water reducing primarily the cold water line fluxes, and (4) pebble drift enhances the hot and warm water reservoir most significantly compared to an equivalent model that only differs in its exclusion of drift (model~C+D+Co).

\section{Discussion}
\label{sec:discus}

\subsection{Contributing to the diversity in water spectra\label{sec:discus_observ}}

\begin{figure*}
    \centering
    \includegraphics[width=1.0\linewidth]{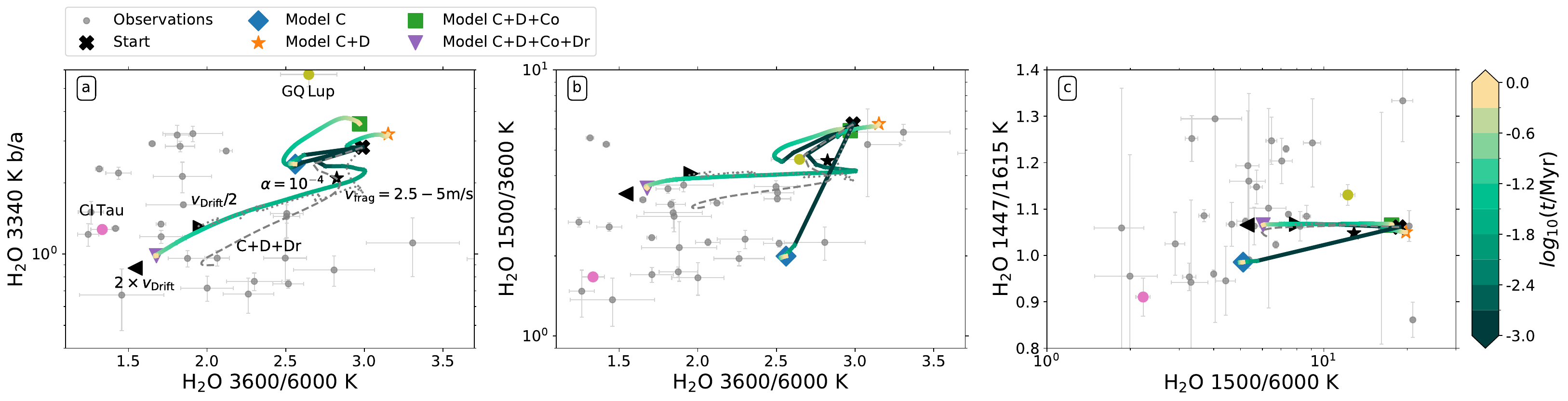}
    \caption{Water line diagnostic diagram for the evolution of the model~C (blue), model~C+D (orange), model~C+D+Co (green), and model~C+D+Co+Dr (purple). CI\,Tau (pink) and GQ\,Lup (olive) are highlighted in all panels. The starting position, with constant water concentrations (for a short discussion on that, see Sect.~\ref{sec:effect_diff}) is marked by a black cross. The two black triangles and the black star show the end position for the supporting models with half/double the fiducial drift velocity, and model~C+D with $\alpha=10^{-4}$ (see Table~\ref{tab:grid}). The evolutionary tracks of model~C+D+Dr (dashed) and model~C+D+Co+Dr with $v_{\mathrm{frag}}=2.5-5\,\rm m/s$ (dotted) are shown in grey. The temperatures indicate the upper level energies \citep{Banzatti2025}. The x-axis of panel~a and b shows the ratio of water lines tracing warm water emission ($> 400 \, \rm K$) compared to hot water ($\sim 850 \, \rm K$) as shown in the upper panels of Fig.~\ref{fig:specs}. The x-axis of panel~c indicates the cold/hot ratio. The y-axis of the panel~a shows the ratio of two lines with different Einstein-A coefficients but similar upper level energies, which are therefore tracing the column densities of $\sim 400\, \rm K$ water. The vertical axis of panel~b shows the ratio between water lines tracing cold emission ($\sim 200\, \rm K$) and warm emission. Panel~c shows the ratio between the two cold water lines on its vertical axis, with the cold/hot ratio shown on the horizontal axis. The grey dots are JWST/MIRI observations \citep{Banzatti2025,Temmink2025}.}
    \label{fig:diagnostics}
\end{figure*}

We compare the simulated spectra of all models to JWST/MIRI observations using the spectra themselves but also the aforementioned line diagnostics introduced by \cite{Banzatti2025}.

The modelled spectra are overplotted with scaled spectra of CI\,Tau \citep{Banzatti2023} and GQ\,Lup \citep{Romero-Mirza2024} to show similarities between simulated and real observations (Fig.~\ref{fig:specs}). Examples of hot-water dominated and cold-water-enriched spectra that have been discussed in detail in literature \citep[e.g.][]{Banzatti2023,Romero-Mirza2024,Banzatti2025,Temmink2025}. We scale their spectra by arbitrary factors to match the peak flux of the $E_{\mathrm{U}}=3645\,\rm K$ water line and focus on the relative strengths of their lines. None of our models attempts to fit any of the observations. 

Model~C shows a good match with the JWST/MIRI spectrum of CI\,Tau. This object was in fact previously reported and used as a template of is an example of a water spectrum dominated by hot water \citep[][]{Banzatti2023,Banzatti2025}. The similarity could be interpreted as a lack of relevance of the other processes in this particular object. This could be achieved by low turbulence (minimising the impact of dynamical processes, see Sect.~\ref{sec:chem_dyna}). Alternatively, higher and/or strongly radially varying photodissociation rates for this object could be balanced by dynamical processes leading to similar spectral features. However, a full exploration of different models to fit observations is beyond the scope of this work. All other models share some spectral features with GQ\,Lup, previously proposed to be a disc dominated by colder water temperatures \citep[][]{Romero-Mirza2024,Banzatti2025}. Especially, the strength of the two cold water lines compared to their neighbouring lines is comparable between GQ\,Lup and model~C+D, model~C+D+Co, and model~C+D+Co+Dr. The closest match in the diagnostic water lines to GQ\,Lup can be seen for model~C+D with $\alpha=10^{-4}$ (panels~a in Fig.~\ref{fig:support_specs}). However, the increased strength of the hot water lines in model~C+D+Co+Dr is not seen in the observation. The mismatch is larger for the high drift model (model~C+D+Co+Dr with $2\times v_{\mathrm{frag}}$) and less for model~C+D+Co+Dr with $v_{\mathrm{frag}}/2$ and model~C+D+Co+Dr with $v_{\mathrm{frag}}=2.5-5\,\rm m/s$ (Fig.~\ref{fig:support_specs}). Interestingly, model~C+D+Dr which shows a higher water concentration in the region above the ice reservoir has weaker cold water lines than GQ\,Lup when normalised to the warm water line. The lack of fragmentation in this model stops the typically seen (e.g. model~C+D+Co+Dr) increase in dust opacity in the inner region. Therefore, even though the cold water lines are strongest compared to all other drift models (Fig.~\ref{fig:support_specs}) the higher temperature water lines are increased even more.

We compare the models to a larger sample of observations using the line diagnostics (Fig.~\ref{fig:diagnostics}). The diagnostics use different ratios between the water lines highlighted in Fig.~\ref{fig:specs} using their integrated flux (as seen in Fig.~\ref{fig:line_strength}). We show three different diagnostic diagrams. The x-axis of the first two (panel~a and b of Fig.~\ref{fig:diagnostics}) uses the ratio between the warm and hot water line. The line ratio on the vertical axis of the panel~a uses two water lines with similar upper level energies but different Einstein A coefficients. This ratio is commonly interpreted as a column density tracer of warm ($T\approx400\,\rm K$) water, with lower ratios corresponding to higher column densities \citep{Banzatti2025}. Panel~b indicates on the vertical axis the ratio between the sum of both cold water lines to the warm water line. This means that the upper half in panel~b is interpreted as a cold water excess. The last diagnostic diagram (panel~c of Fig.~\ref{fig:diagnostics}) shows the cold/hot ratio on the horizontal and the ratio between the two cold water lines on the vertical axis. Higher ratios between the two cold water lines are interpreted as an excess in the coldest observable water ($<200\,\rm K$).

The first conclusion from the model comparison to observed line diagnostics is that the evolutionary tracks strongly overlap with the observations increasing our confidence on the realism of our simulations. Additionally, it becomes clear that the evolutionary processes can lead to drastic changes in the line ratios highlighting the need to account for the interplay of all of them when analysing observations. However, some observed ratios are not reproduced by any model track. None of our models significantly increases the coldest water ratio (y-axis of panel~c of Fig.~\ref{fig:diagnostics}). This is not further surprising since all models use the same stellar and structural parameters. Similarly, all models (except for model~C) show cold/warm water ratios that are typically associated with an cold water excess. This can be a consequence of the disc setup (e.g. the initialisation with constant water concentrations and the assumed photodissociation rate). The impact of different disc setups (including the effect of different dust opacities and dust-to-gas ratios) will be tested in future work.

\subsubsection{Impact of chemical processing on line diagnostics\label{sec:disc_impact_C}}

Next, we focus on the individual tracks of all models. 
For model~C, both the cold/warm and warm/hot lines (blue marker in panel~b of Fig.~\ref{fig:diagnostics}) decrease strongly over time (from $6.2$ and $3.0$ to $2.0$ and $2.5$, respectively). Interestingly, the reduction in column density at $T<400\,\rm K$ for model~C does not lead to an increase in the column density tracer (y-axis of panel~a) but to a small decrease (from $\sim2.7$ to $\sim2.3$). This is due to an optical depth effect: emission at higher temperatures is optically thicker given the same column density. If the contribution from low temperatures to these lines is strongly reduced during the model's evolution, a larger fraction of the emission becomes optically thick, which results in an increase in the column density tracer ratio. This example illustrates how column density reductions can change this tracer in a way that is typically associated with column density increases. 
Panel~c of Fig.~\ref{fig:diagnostics} shows the ratio of the two cold water lines compared to the cold/hot radio. The cold/hot ratio decreases for model~C (from $\sim19$ to $\sim 5$), with the ratio of the two cold lines decreasing slightly as well (from $\sim1.05$ to about $1.0$). Model~C occupies a similar diagnostic region as CI\,Tau in all cold water tracers, but CI\,Tau's column density tracer indicated higher column densities and an even stronger trend towards higher temperatures in all temperature tracers.

\subsubsection{Impact of diffusion and advection on line diagnostics\label{sec:diagnositc_diff}}

The water line diagnostics for model~C+D show a fast drop compared to the initial conditions by $0.6$ and $0.1$ in the cold/warm and warm/hot tracer, respectively. This is followed by a long term increase of $0.6$ and $0.2$, respectively. Therefore, model~C+D moves slightly to the top right in panel~b of Fig.~\ref{fig:diagnostics}, which is typically associated with a cold water excess due to pebble drift. This shows that water diffusion into the cold water reservoir above the snow region can lead to a similar effect as often associated with pebble drift. We note that the diagnostic region occupied by model~C+D at $1\,\rm Myr$ corresponds to a stronger enhancement of colder temperature water lines than seen in most observation (even observations with a cold water excess) potentially due to the assumed photodissociation rate. As shown in Sect.~\ref{sec:chem_dyna}, the strength of this replenishment also depends on the disc turbulence. Therefore, we predict in this scenario an observable relation between mass accretion rate  as a proxy for disc turbulence and the cold water line flux. It can for example be seen that model~C+D with $\alpha=10^{-4}$ occupies a position between model~C+D and model~C just as expected. The column density tracer (panel~a) rises by about $0.3$ due to the dominance of optically thinner colder emission. Lastly, the coldest water line ratio does not change significantly over time (within $0.02$) but is significnalty higher than the final ratio of model~C. All temperature tracers show strong overlap between model~C+D and GQ\,Lup with the column density in GQ\,Lup being slightly lower than values extracted from our models.

\subsubsection{Impact of non-drifting pebbles on line diagnostics}

Model~C+D+Co behaves similarly to model~C+D in the line diagnostic space with, for example, no strong change in its coldest water ratio over time, which corresponds to a higher ratio than model~C at $1\,\rm Myr$. The column density tracer decreases first to $~2.4$ due to same effect as discussed for model~C, with a subsequent increase to $3.6$ due to the redistribution of water to lower temperatures (see Section~\ref{sec:results_no_drift}). The warm/hot ratio decreases to $2.5$ before rising again to its initial value. Similarly, the cold/warm ratio decreases to $4.6$ before reaching $5.8$ at $1\,\rm Myr$. This means that the cold water enhancement seen in panel~b is slightly less extreme than the one seen in model~C+D, but still comparable to the strongest cold water excess observations.

\subsubsection{Impact of pebble drift on line diagnostics}

The line diagnostic ratios of model~C+D+Co+Dr, the model including pebble drift, evolve over time towards higher temperatures. The track leads to lower values of cold/warm ($3.6$) and warm/hot ($1.7$) line ratios at $1\,\rm Myr$ compared to the initial conditions, model~C+D, and model~C+D+Co. This is surprising since typically observations with a high cold/warm and warm/hot ratio are associated with pebble drift dominated discs \citep[e.g.][]{Banzatti2023,Romero-Mirza2024,Banzatti2025}. However, as discussed in Sect.~\ref{sec:diagnositc_diff} the cold/warm water ratio seen in model~C+D is high compared to cold water excess observations (potentially due to the assumed photodissociation rate). Therefore, even the reduction in this ratio by model~C+D+Co+Dr leaves the model comparable with observations associated with domination of colder water temperatures. This means that cold water excess observations are compatible with models including pebble drifts even though the effect of drift compared to identical models without drift acts by reducing the cold/warm water ratio. The coldest water tracer (panel~c) only increase by a small fraction. From the small grid of models, we cannot conclude if different model setups including drift could lead to a more significant change. Generally, the coldest and cold/warm diagnostics of model~C+D+Co+Dr are similar to GQ\,Lup. However, all ratios involving the hot water line are significantly shifted to higher temperatures in the model. Therefore, these tracers are more similar to CI\,Tau. The column density diagnostic (panel~a of Fig.~\ref{fig:diagnostics}) shows a strong decrease (to $~1.0$) typically associated with higher column densities. This is in line with the strong increase of column density at $T>400\,\rm K$. Additionally, we note that the models with higher and lower drift velocity (see Table~\ref{tab:grid}) follow very similar evolutionary diagnostic tracks. The model with doubled the drift velocity continues the path set by model~C+D+Co+Dr but continues e.g. to cold/warm and warm/hot ratios of $3.36$ and $1.56$, respectively. The model with half the drift velocity on the other hands reaches values ($4.04$ and $1.94$) that are already surpassed earlier by model~C+D+Co+Dr. Similarly, increasing pebble fluxes lead to lower cold/hot water ratios contrary other predictions \citep[e.g. Eq.~3 and Eq.~4 in ][]{Krijt2025}. The fundamental difference arises from the previously made assumption that pebble drift affects the observable cold water mass while higher temperatures are unaffected \citep[e.g. Eq.~11 in][]{Romero-Mirza2024}. However, comparing the observable water mass for three models including pebble drift at $1\,\rm Myr$, we find that the cold water mass, which we define as all emission at $T<400\,\rm K$, varies only by $\sim10\,\%$ with model~C+D+Co+Dr with $v_{\mathrm{Drift}}/2$ actually having the highest observable cold water mass ($\sim5.77\times 10^{-6}\,\rm M_\oplus$) and model~C+D+Co+Dr with $2\times v_{\mathrm{Drift}}$ having the lowest ($\sim5.22\times 10^{-6}\,\rm M_\oplus$). This difference is negligible when considering the observable water mass at $T>400\,\rm K$ increasing from $\sim3.35\times 10^{-6}\,\rm M_\oplus$ (for model~C+D+Co+Dr with $v_{\mathrm{Drift}}/2$) to $\sim14.4\times 10^{-6}\,\rm M_\oplus$ (for model~C+D+Co+Dr with $2\times v_{\mathrm{Drift}}$).

Model~C+D+Co+Dr with $v_{\mathrm{frag}}=2.5-5\,\rm m/s$ roughly follows the diagnostic traces of model~C+D+Co+Dr. This shows that a fragmentation velocity change within a factor of $2$ does not have significant impact of the observable diagnostics even though the underlying water and dust density structure differs. Panel~d of Fig.~\ref{fig:support_conditions} shows the emitting conditions of this model, which are notably different from model~C+D+Co+Dr. The observed column density peaks sharply at about $400-500\,\rm K$ with column densities at higher temperatures being lower than in model~C+D+Co+Dr. The temperature-confined increase comes from the fact that the change in fragmentation velocity leads to a change in pebble size and therefore drift speed. This means that pebbles sublimate their ices in a traffic jam at the transition between both regimes. Since the column density peaks moves to higher temperatures over time, the diagnostics move to higher temperature regions. Additionally, the $\tau_{\rm IR}=1$ line increases inside the water snowline hiding more water vapour as already discussed by \cite{houge_smuggling_2025}. Model~C+D+Dr which has as the only models shown efficient outwards water vapour diffusion into the region above the ice reservoir shows lower cold/warm ratio than model~C+D+Co+Dr ($3.0$ instead of $3.6$). This is due to the strong increase in the warm (and hot) water line due to the lack of increasing dust opacity in the inner disc (as discussed in Sect.~\ref{sec:discus_observ}). This highlights the role of dust opacity when interpreting mid-IR water spectra. The diffusion of water from the vertically mixed plume occurs outwards leading to an decrease in the coldest water line ratio ($1.03$) compared to model~C+D+Co+Dr ($1.05$). Therefore, an increase in the coldest water line ratio requires additional effects than sublimating ice in the midplane that diffuses into the coldest observable regions. We discuss these aspects in Sect.~\ref{sec:possible_solutions}.

\subsection{Emission regions of diagnostic water lines \label{sec:discus_cold_enhancement}}

In this section, we link the emission of the diagnostic water lines to their emitting regions. 
We calculate the radial ranges from which $15\,\%$ to $85\,\%$ of the radially cumulative flux originates for the hot, warm, and cold water lines. With additional predictions for three far-IR lines at $40.69\,\rm \mu m$ ($E_{\rm U}=550\,\rm K$), at $46.48\,\rm \mu m$ ($E_{\rm U}=410\,\rm K$), and at $61.81\,\rm \mu m$ ($E_{\rm U}=552\,\rm K$) which will be observed by future far-IR mission like the Planetary Origins and Evolution Multispectral Monochromator (POEMM\footnote{https://poemm.astro.cornell.edu}) and PRobe far-Infrared Mission for Astrophysics \citep[PRIMA\footnote{https://prima.ipac.caltech.edu};][]{Moullet2025}. These radial ranges are plotted on top of all water concentration panels in Fig.~\ref{fig:pebble_compare_alpha} and the comparison panels in Fig.~\ref{fig:pebble_full_2d}. 

Most of the flux of the JWST/MIRI cold water lines comes from the water reservoir above the ice region. This has been shown using full thermochemical models by \cite{Vlasblom2025}. We can confirm this behaviour even if diffusion (model~C+D; panel~a of Fig.~\ref{fig:pebble_compare_alpha}), diffusion with $\alpha=10^{-4}$ (panel~b of Fig.~\ref{fig:pebble_compare_alpha}), additional coagulation and fragmentation and pebble drift (model~C+D+Co+Dr; panels~4 of Fig.~\ref{fig:pebble_full_2d}) are accounted for.

However, both the warm and cold water lines overlap radially with the midplane snowline in all cases (panel~a and panel~b of Fig.~\ref{fig:pebble_compare_alpha} and panels~4 of Fig.~\ref{fig:pebble_full_2d}). The emission regions are shifted to smaller radii and higher temperatures with a stronger radial concentration gradient as seen most prominently in model~C+D+Co+Dr (panel~4d of Fig.~\ref{fig:pebble_full_2d}). 
Therefore, when pebble drift enhances the water concentration inside the midplane snowline ($>400\,\rm K$ in the disc surface), the cold water lines are getting stronger (see e.g. panel~2d of Fig.~\ref{fig:specs}). However, future far-IR observatories like POEMM and PRIMA will probe colder water lines and truly probe the coldest water emission $<400\,\rm K$.

\subsection{Limitations and future work \label{sec:possible_solutions}}

This study shows that the link between cold water line excess and pebble drift is not as straight forward as previously thought. However, given the observed tentative correlation between drift dominated discs and cold water excess \citep[e.g.][]{Banzatti2023,Romero-Mirza2024,Krijt2025} we explore different scenarios that could lead to that correlation, in preparation for a larger parameter study that is planed for future work.

As shown in the Sect.~\ref{sec:results_drift}, the cold water emitting region above the snow surface could be enhanced by drift if outwards transport in the disc surface is efficient. We explain the lack of outwards diffusion in model~C+D+Co+Dr by a water cycle including vertical diffusion, freeze out, coagulation and inwards pebble drift. This cycle has been identified as well using 2D multifluid hydrodynamic simulations by \cite{Wang2025} who highlighted the dependence of the cycle on the snowline shape and radial temperature gradient. The efficiency of this cycle depends on all its individual processes.

Stronger advection fields \citep[e.g.][]{Ciesla2009} can under the right conditions increase the cold water reservoir.
Similarly, a change of the ratio between radial and vertical $\alpha$-parameter can lead to the same effect. Constant (radial and vertical) disc turbulence is a simplifying assumption that has been challenged by multiple studies \citep[e.g.][]{Turner2014,Rosotti2023}. Tuning the ratio between the vertical and radial $\alpha$ could lead to stronger radial diffusion which could help increasing the cold water reservoir concentration.

Alternately, an important factor that determines which water reservoir is enhanced strongest through drift is the temperature difference between the midplane and disc surface. Our models make some necessary simplifying assumption. We assume the same temperature structure for dust and gas, while several studies show that the gas and dust temperature can differ significantly \citep[e.g.][]{Woitke2009}. Additionally, we ignore temperature evolution, which given the changes in dust densities in model~C+D+Co+Dr can be significant. We also omitted accretion heating in the disc midplane, which can lower the midplane surface temperature difference \citep{Calahan2026}, pushing out the midplane snowline, and therefore lead to a more direct correlation between observable cold water fluxes and pebble drift. Similarly, \cite{Wang2025} show that latent heat exchange during ice sublimation can flatten the temperature gradient across the snowline.

While we assume a constant pebble flux in this study, 1D models predict strong variations in pebble fluxes within $1\,\rm Myr$ \citep[e.g.][]{Birnstiel2012,Drazkoska2021}. Future works will study the influence of these profiles on JWST/MIRI observables using our 2D modelling approach.

Additionally, we simplify the water chemistry compared to full thermo-chemical models (e.g. assuming a constant photodissociation timescale of $40\,\rm yr$). This enables us to evolve 2D dynamics and chemistry together. Combining thermo-chemical models with abundances based on 1D evolutionary codes can provide further insight into the interplay between detailed chemistry (including temperature changes) and dynamics (Vlasblom et al. submitted).

A larger parameter study can be beneficial to assess if the observed cold water assess in some discs could be explained by disc conditions independent of pebble flux. Similarly, future modelling efforts should incorporate the most significant phenomena mentioned above. Additionally, 1D evolution models predict that several molecular ratios strongly evolve with time \citep[e.g.][]{Sellek2025}. This shows the benefit of not just using water, but also other molecules to assess the impact of pebble drift on JWST/MIRI observations. 

From an observational point of view, far-IR missions like POEMM and PRIMA will provide a unique opportunity to probe water lines emitted at $T<400\,\rm K$, with high spectral resolution that resolves the lines and allows for a determination of their radial emission region.

\section{Conclusions}
\label{sec:conclusion}

In this study, we introduce the MAGPIE model (Sect.~\ref{sec:method}) to analyse the effect of dynamical and chemical processes on the 2D water concentration structure (Sect.~\ref{sec:results_density}) and the water emission observed with JWST/MIRI (Sect.~\ref{sec:results_spec}). For a single disc set-up, we ran a small grid of models (Table~\ref{tab:grid}) including different processes to examine the impact of them on the 2D water density structure and simulated JWST/MIRI spectra. The main conclusions are as following:

\begin{enumerate}
    \item For a single disc set-up, including/excluding transport processes leads to changes in simulated JWST/MIRI spectra (Fig.~\ref{fig:specs}) and in most diagnostic line rations (Fig.~\ref{fig:diagnostics}) that are comparable to the spread currently seen in the observed population (Sect.~\ref{sec:discus_observ}).
    \item Transport, in the form of diffusion, advection, and pebble drift compared to a static chemical model affects all traced concentrations in every modelled grid cell (Sect.~\ref{sec:results_density}). Therefore, dynamics cannot be ignored when describing the inner few au of a protoplanetary disc.
    \item Starting from constant abundances, chemistry mainly influences the observable water reservoir by photodissociating the cold water surface layer (Sect.~\ref{sec:results_chem}) consistent with results by \cite{Vlasblom2025}. Chemical water formation enhances cold water in the photodissociated region to abundances of $\sim10^{-6}$ (Sect.~\ref{sec:results_chem}). While our models only account for interstellar UV, photodissociation due to stellar UV can significantly lower this abundance. 
    \item Turbulent diffusion reduces concentration gradients in the disc leading to an increase of water vapour in the photodissociation region compared to a model without diffusion (Sect.~\ref{sec:chem_dyna}). Therefore, diffusion can lead to an (turbulence strength dependent) increase in lower temperature water tracers (Sect.~\ref{sec:discus_observ}) which for the assumed photodissociation rate are extreme cold/warm ratios compared to most cold water excess observations. However, these models do not match the observed $1447\,\rm K/1615\,\rm K$ line ratio of objects showing a cold water excess which is sensitive to the coldest water reservoir ($T\lesssim200\,\rm K$).
    \item Freeze out onto pebbles at the midplane reduces the observable cold water reservoir via the vertical cold-finger effect (Sect.~\ref{sec:results_no_drift}). 
    \item Drift and sublimation of icy pebbles does not lead to an exclusive enhancement of the observable cold water fluxes compared to an identical model without drifting pebbles. Instead, depending on the vertical temperature profile, sublimating ices can primarily enhance the hot and warm water fluxes, which overshadows the mild increases in the cold water flux (Sect.~\ref{sec:results_spec}). Therefore, pebble drift can counter the extreme cold temperature enhancement seen in models including chemistry and diffusion (Sect.~\ref{sec:discus_observ}).
    \item Radial diffusion is relatively inefficient at distributing the plume of water vapour released by sublimating pebbles to the region responsible for cold water emission above the snow surface. This is due to a conveyor belt effect of freeze out, coagulation, and pebble drift. However, even if this cycle is interrupted (e.g. due to slow coagulation) and outwards diffusion is more efficient, the colder water tracers do not increase compared to our standard drift model due to generally lower dust concentrations which leads to a lower $\tau_{\rm IR}=1$ layer also at small radii. Therefore, larger water reservoirs are observable at higher temperatures (Sect.~\ref{sec:results_drift}). 
    \item A decrease in fragmentation velocity between wet and dry pebbles (of a factor of $2$) increases the dust opacity inside the midplane water snowline. However, in our models even this increase does not lead to diagnostics shifts to lower temperatures (Sect.~\ref{sec:discus_observ}).
    
\end{enumerate}

This study shows the impact of the interplay between chemical and dynamical processes on the water distribution in a protoplanetary disc model. While including different processes can explain some observational different between objects, an extensive parameter study is needed to evaluate the impact of disc structures more generally. This includes not just matching observed line ratios that are currently not reproduced by our models (e.g. high ratios between the coldest water lines) but also comparing to typical retrieved column density and temperature ranges.

\section*{Acknowledgements}

TK was supported by Science and Technology facilities Council (STFC) grant no. ST/Y002415/1. SK was partly supported by Science and Technology facilities Council (STFC) grant no. ST/Y002415/1. JW is funded by the UK Science and Technology Facilities Council (STFC), grant code ST/Y509383/1.

\section*{Data Availability}

The data generated during this study will be shared on reasonable request to the corresponding author.



\bibliographystyle{mnras}
\bibliography{lib} 




\appendix

\section{Additional chemistry information}

\label{sec:h3plus}

\begin{figure}
    \centering
    \includegraphics[width=\linewidth]{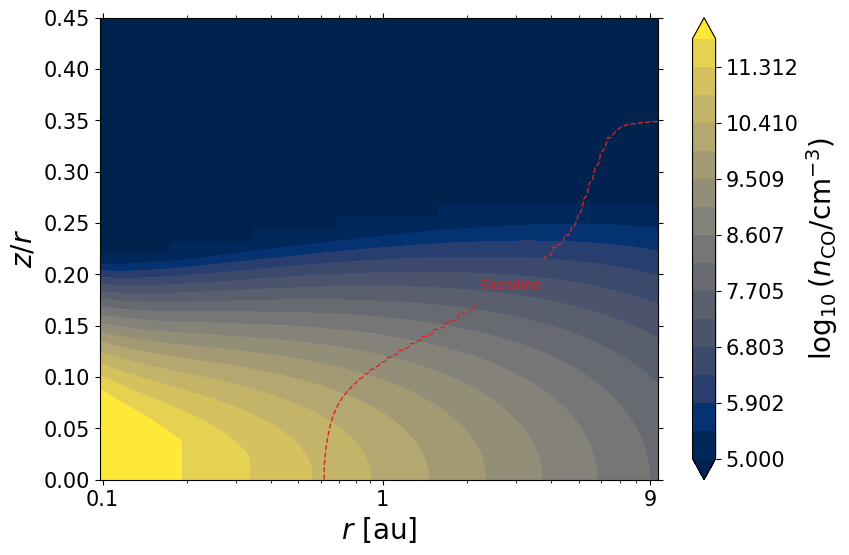}
    \caption{Number density of CO. The colors indicate the 2D number densities with the red line denoting the initial water snowline.}
    \label{fig:num_co}
\end{figure}

\begin{figure}
    \centering
    \includegraphics[width=\linewidth]{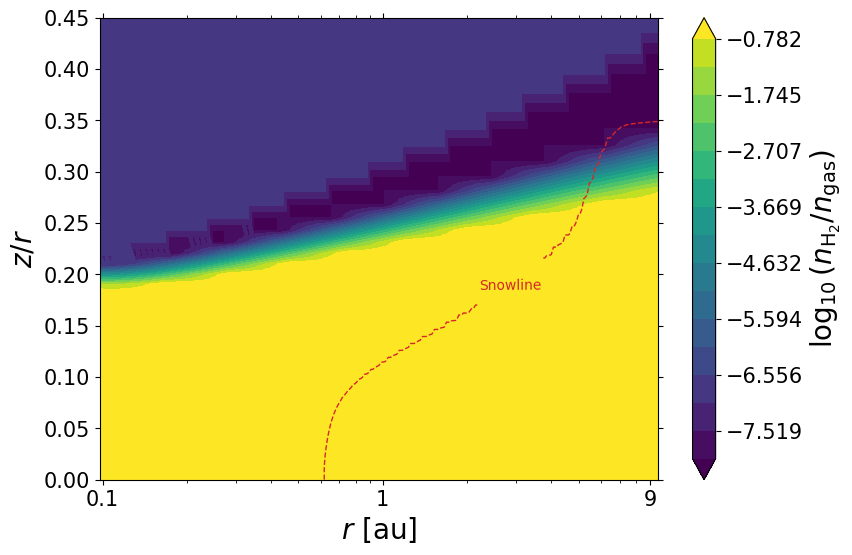}
    \caption{Abundance of \ce{H2}. The colors indicate the 2D abundances with the red line denoting the initial water snowline.}
    \label{fig:abund_h2}
\end{figure}

\begin{figure}
    \centering
    \includegraphics[width=\linewidth]{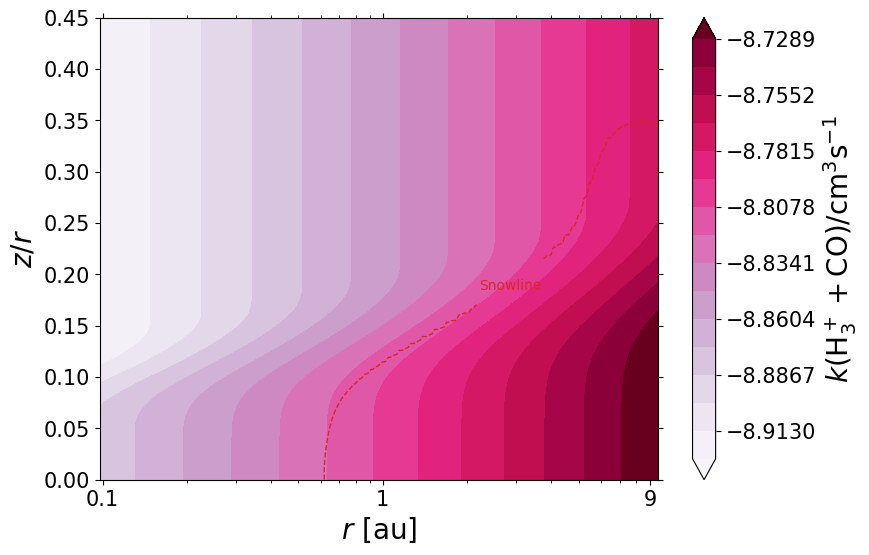}
    \includegraphics[width=\linewidth]{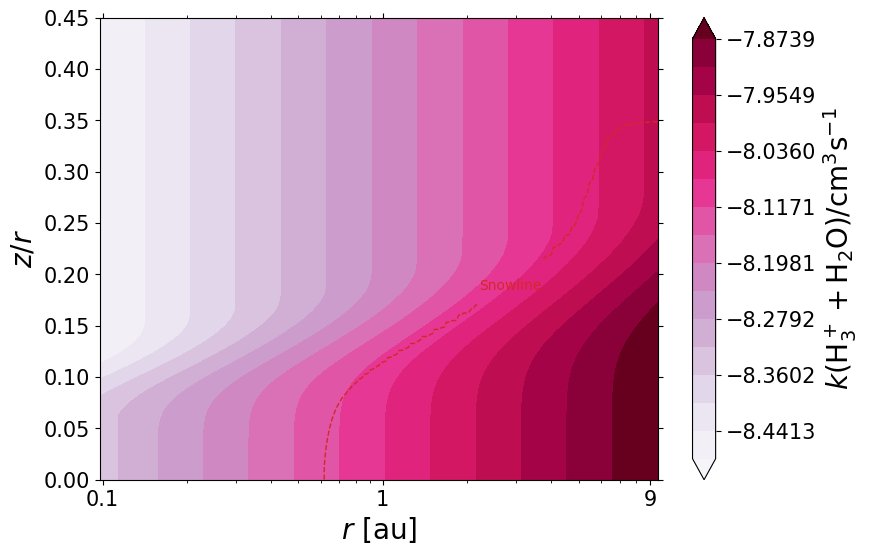}
    \caption{Reaction rates of reaction~\ref{rec:co} (top panel; between \ce{H3+} and \ce{CO}) and reaction~\ref{rec:h2o} (bottom panel; between \ce{H3+} and \ce{H2O}).}
    \label{fig:water_h3plus_react}
\end{figure}

This section contains additional figures explaining the chemical model and a subsection focused on the abundance of \ce{H3+}. Fig.~\ref{fig:num_co} and Fig~\ref{fig:abund_h2} display the number density and abundance of \ce{CO} and \ce{H2}, respectively. The reaction rates of reaction~\ref{rec:co} (between \ce{H3+} and \ce{CO}) and reaction~\ref{rec:h2o} (between \ce{H3+} and \ce{H2O}) are shown in Fig.~\ref{fig:water_h3plus_react}. 


The number density of \ce{H3+} is described in reaction~\ref{eq:nh3+_computation} (repeated below for clarity) by a simple balance between the formation via \ce{H2} and the destruction by \ce{CO} and \ce{H2O}. In this section, we highlighting a few consequences of this reaction equation. As seen in Fig.~\ref{fig:water_h3plus_react}, the reaction rates for the two destruction reactions vary by no more than an order of magnitude within the model setup. Therefore, the number densities of \ce{CO} and water which vary by several orders of magnitude will dominate the amount of \ce{H3+} destruction. 

Below we repeat Eq.~\ref{eq:nh3+_computation} for clarity purposes.

\begin{align}
    n\left(\ce{H3+}\right)&=\frac{\zeta_{\rm CR} n\left(\ce{H2}\right)}{n\left(\ce{CO}\right)k_1+n\left(\ce{H2O}\right)k_2} \ .
\end{align}

\begin{figure}
    \centering
    \includegraphics[width=0.95\linewidth]{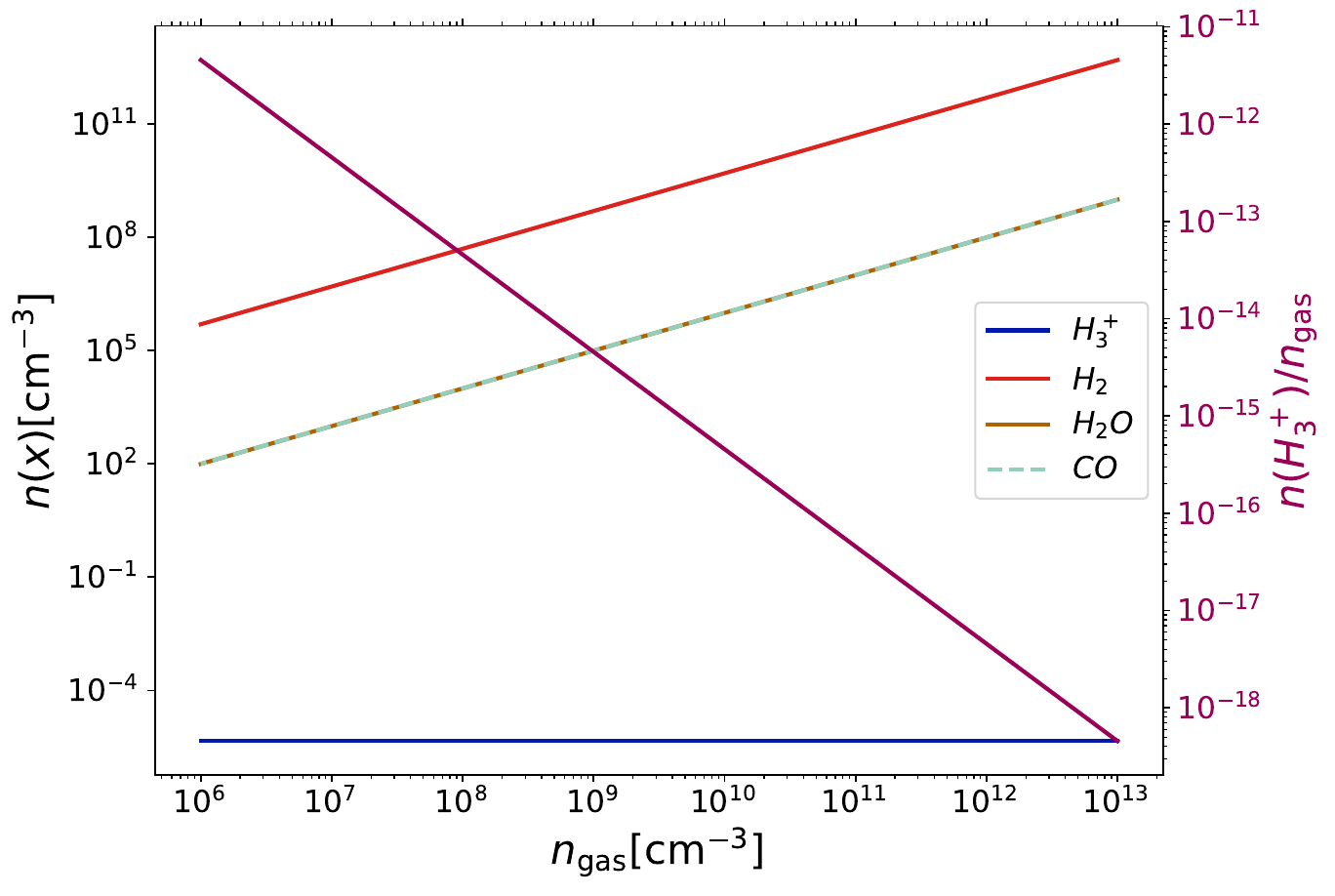}
    \caption{Toy model number densities of \ce{H3+}, \ce{H2}, \ce{H2O}, and \ce{CO} as a function of gas density. The \ce{H3+} abundance (purple) is shown on the right axis.}
    \label{fig:toy_h3plus}
\end{figure}

For the toy model in this appendix, we assume $k_1\approx10^{-9}\,\rm cm^3/s$, $k_2\approx10^{-8}\,\rm cm^3/s$, and $\zeta \approx 10^{-17}\,\rm s^{-1}$. The interesting behaviour of \ce{H3+} comes when examining the relation between molecular abundances and number densities. Assuming a constant abundance of \ce{H2}, \ce{H2O}, and \ce{CO} (for this model $0.5$, $10^{-4}$, and $10^{-4}$, respectively) an increase in the gas number density will be reflected by an equivalent increase in all molecular number densities (see Fig.~\ref{fig:toy_h3plus}). However, due to the nature of Eq.~\ref{eq:nh3+_computation} equivalent increases in \ce{H2} will be cancelled out by \ce{CO} and \ce{H2O} leading to a constant number density of \ce{H3+} which depends on the assumed molecular abundances, ionization rate, and reaction rate \citep[as already shown by][]{Lepp1987}. Therefore, the \ce{H3+} abundance will increase with decreasing gas number density, making gas-phase water formation via \ce{H3+} more prevalent in lower mass discs if all other factors are assumed to be constant.

\section{Support model figures}
\label{sec:support}

\begin{figure*}
    \centering
    \includegraphics[width=0.95\linewidth]{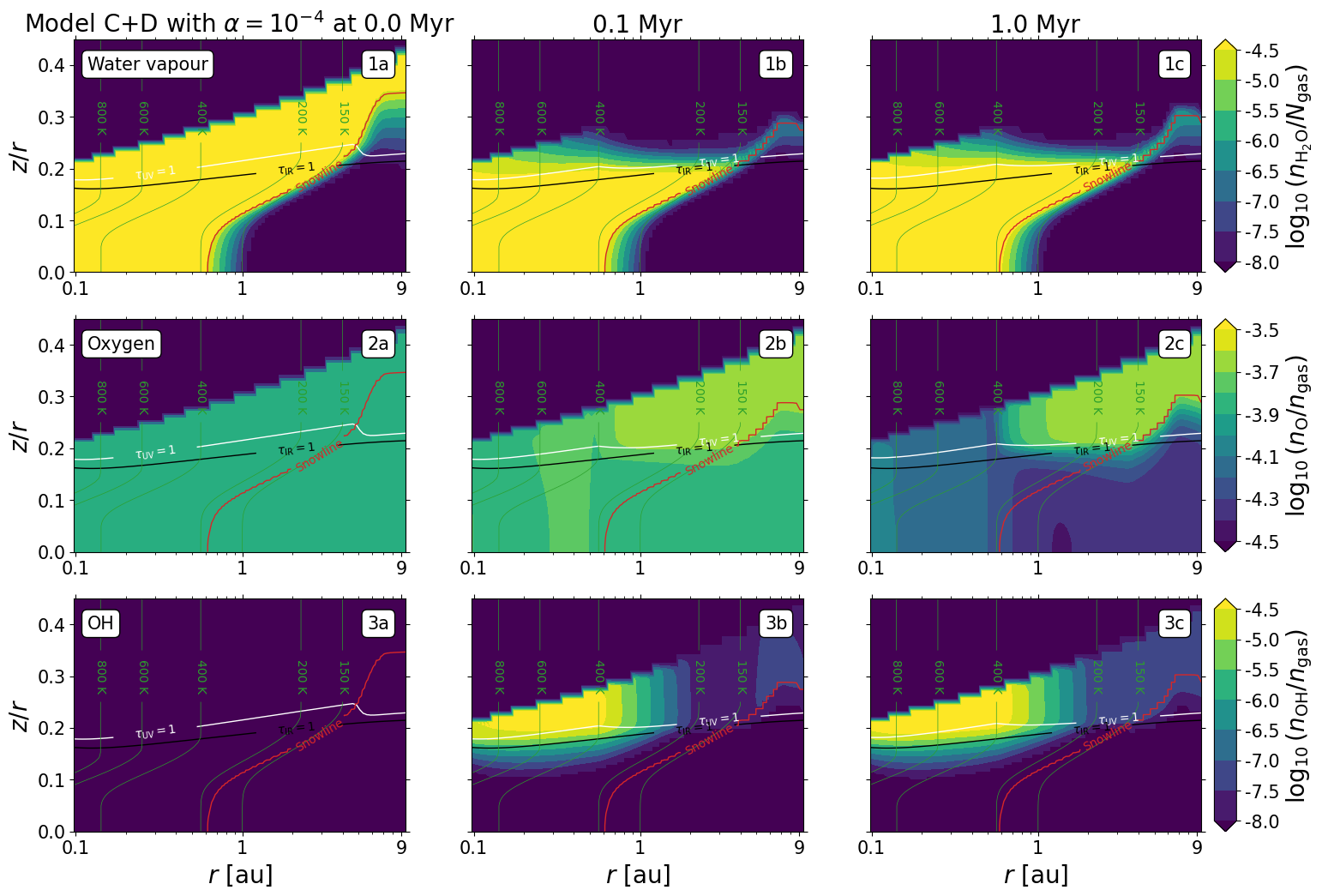}
    \caption{The chemical and dynamical evolution of water in model~C+D with $\alpha=10^{-4}$ (first supporting model in Table~\ref{tab:grid}). Every column depicts a different time during the evolution (initial conditions, $0.1\,\rm Myr$, and $1\,\rm Myr$) of the water abundance (top row), oxygen abundance (second row), and \ce{OH} abundance (last row). The snowline, temperature contours and opacity lines are shown for reference.}
    \label{fig:support_chem_and_dyna_2d}
\end{figure*}

\begin{figure*}
    \centering
    \includegraphics[width=1.0\linewidth]{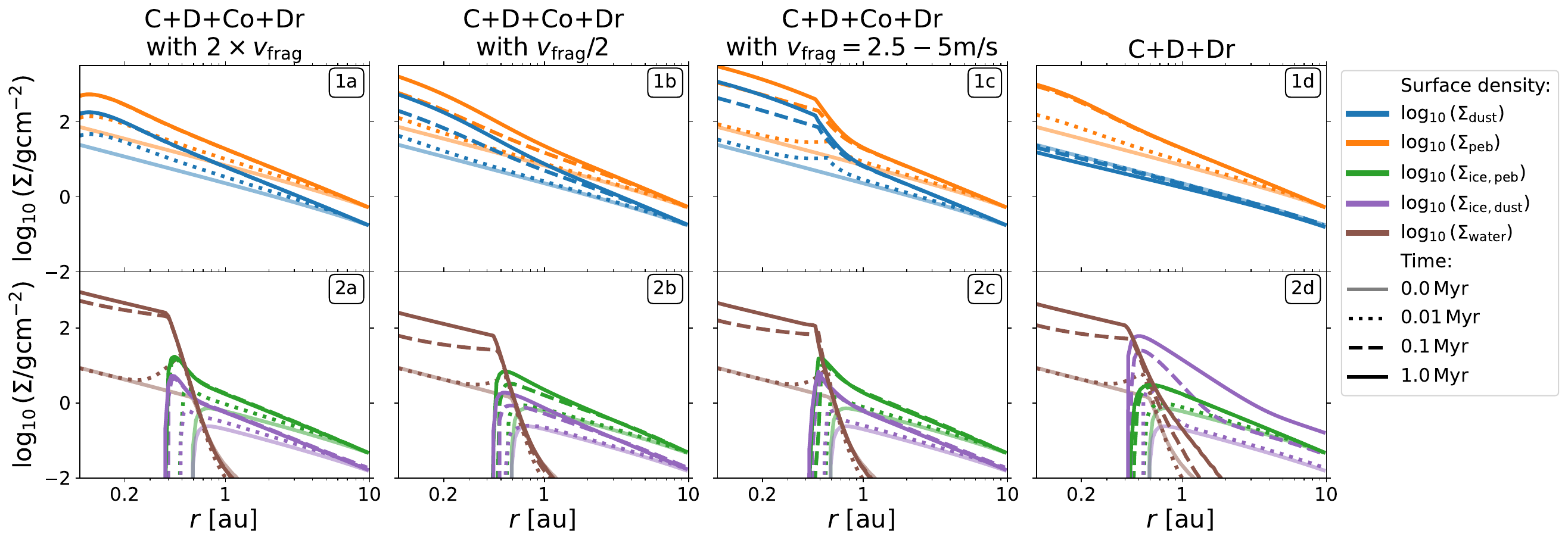}
        \caption{Evolution of surface densities for model~C+D+Co+Dr with $2\times v_{\mathrm{frag}}$ (panels~a), model~C+D+Co+Dr with $v_{\mathrm{frag}}/2$ (panels~b),  model~C+D+Co+Dr with $v_{\mathrm{frag}}=2.5-5\,\rm m/s$ (panels~c), and model~C+D+Dr (panels~d). The different colours correspond to different surface densities with the dust, pebble, ice on pebbles, ice on dust grains, and water vapour surface density depicted in blue, orange, green, purple, and brown, respectively. The upper panels show all solids while the lower panels depict all water surface densities. The initial conditions (faint lines) are the same for all models. The surface densities at $0.01\,\rm Myr$ (dotted line), $0.1\,\rm Myr$ (dashed line) and $1\,\rm Myr$ (solid line) are displayed as well.}
    \label{fig:support_surface_dens}
\end{figure*}

\begin{figure*}
    \centering
    \includegraphics[width=1.0\linewidth]{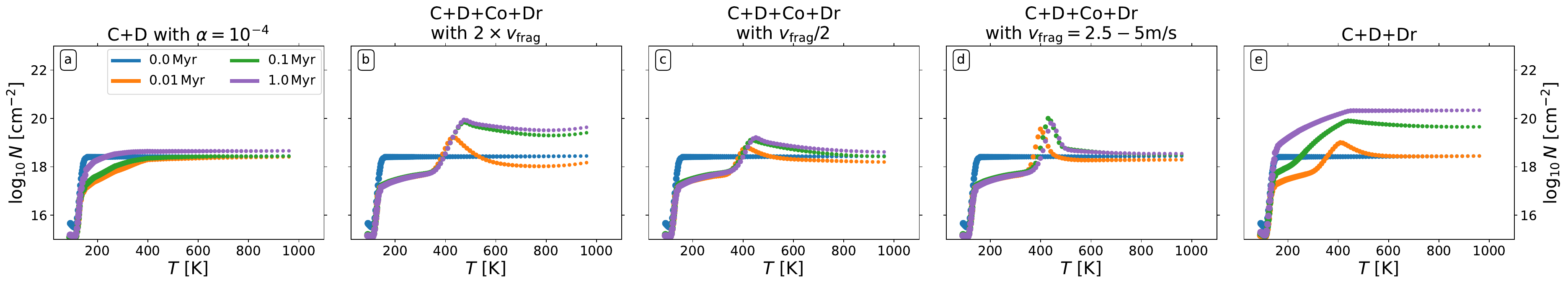}
    \caption{Slab conditions extracted for model~C+D with $\alpha=10^{-4}$ (panel~a), model~C+D+Co+Dr with $2\times v_{\mathrm{frag}}$ (panel~b), model~C+D+Co+Dr with $v_{\mathrm{frag}}/2$ (panels~c),  model~C+D+Co+Dr with $v_{\mathrm{frag}}=2.5-5\,\rm m/s$ (panel~d), and model~C+D+Dr (panel~e) at $0.0\,\rm Myr$ (blue), $0.01\,\rm Myr$ (orange), $0.1\,\rm Myr$ (green), and $1.0\,\rm Myr$ (purple). Every marker represents a single column of the 2D density structure turned into input for a slab model (column density above $\tau_{\mathrm{IR}}=1$ and mass average temperature in that visible column). The marker size is proportional to the logarithm of the slab's emitting area.}
    \label{fig:support_conditions}
\end{figure*}

\begin{figure*}
    \centering
    \includegraphics[width=\linewidth]{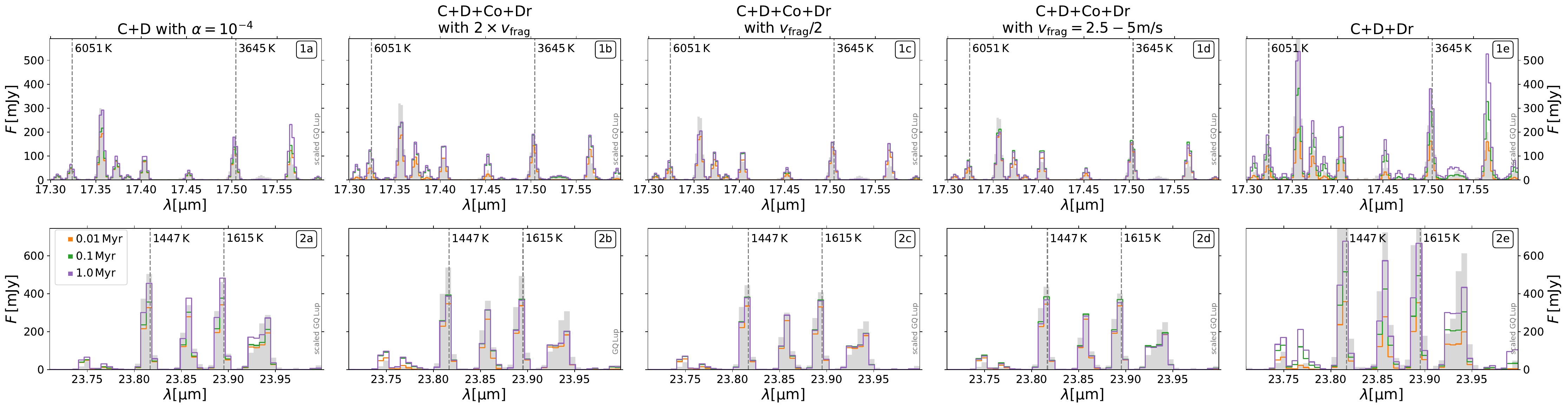}
    \caption{Spectral zoom-ins for model~C+D with $\alpha=10^{-4}$ (panels~a), model~C+D+Co+Dr with $2\times v_{\mathrm{frag}}$ (panels~b), model~C+D+Co+Dr with $v_{\mathrm{frag}}/2$ (panels~c),  model~C+D+Co+Dr with $v_{\mathrm{frag}}=2.5-5\,\rm m/s$ (panels~d), and model~C+D+Dr (panels~e) at $0.01\,\rm Myr$ (orange), $0.1\,\rm Myr$ (green), and $1.0\,\rm Myr$ (purple). Overplotted is the JWST/MIRI spectrum of GQ\,Lup scaled to the peak flux of the $E_{\mathrm{U}}=3645\,\rm K$ water line of the $1\,\rm Myr$ model for comparison (details in Sect.~\ref{sec:discus_observ}). The zoom-in from $17.29\,\rm \mu m$ to $17.6\,\rm \mu m$ (top row) contains lines that are used to analyse hot water ($\sim 17.32\,\rm \mu m$) and warm water ($\sim 17.5\,\rm \mu m$) marked with their upper level energy. The zoom-in from $23.71\,\rm \mu m$ to $24\,\rm \mu m$ (bottom row) contains two marked cold water lines.}
    \label{fig:support_specs}
\end{figure*}

In this section, we present additional figures regarding the supporting models. This includes the same 2D abundance figure as shown for model~C+D (Fig.~\ref{fig:chem_and_dyna_2d} ) but for $\alpha=10^{-4}$ (Fig.~\ref{fig:support_chem_and_dyna_2d}), the surface density evolution of all supporting models including drift (Fig.~\ref{fig:support_surface_dens}), the emitting conditions extracted from all supporting models (Fig.~\ref{fig:support_conditions}) and the resulting spectra (Fig.~\ref{fig:support_specs}).

\bsp	
\label{lastpage}
\end{document}